\documentclass[universe,review,accept,oneauthor]{Definitions/mdpi} 
\firstpage{1} 
\pubvolume{11}
\issuenum{8}
\articlenumber{270}
\pubyear{2025}
\copyrightyear{2025}
\externaleditor{Gerald B. Cleaver} 
\datereceived{26 June 2025} 
\daterevised{26 July 2025} 
\dateaccepted{30 July 2025} 
\datepublished{15 August 2025} 
\hreflink{https://doi.org/10.3390/\\universe11080270} 

\usepackage{tikz}

\Title{Sculpting
 Spacetime: Thin Shells in Wormhole Physics}

\TitleCitation{Sculpting Spacetime: Thin Shells in Wormhole Physics}

\Author{Francisco S. N. Lobo 
 $^{1,2}$}
 \AuthorCitation{Lobo, F.S.N.}
\address{%
$^{1}$ \quad Instituto de Astrof\'{i}sica e Ci\^{e}ncias do Espa\c{c}o, Faculdade de Ci\^encias da Universidade de Lisboa, Edif\'{i}cio C8, Campo Grande, P-1749-016 
 Lisbon, Portugal; fslobo@fc.ul.pt\\
$^{2}$ \quad Departamento de F\'{i}sica, Faculdade de Ci\^{e}ncias, Universidade de Lisboa, Edifício C8, Campo Grande, P-1749-016 Lisbon, Portugal}

\abstract{
	In this work, we employ the Darmois--Israel thin-shell formalism to construct both static and dynamic thin-shell configurations surrounding traversable wormholes. Initially, using the cut-and-paste technique, we perform a linearized stability analysis in the presence of a general cosmological constant. Our results show that for sufficiently large positive values of the cosmological constant—corresponding to the Schwarzschild–de Sitter geometry—the stability regions of the wormhole solutions are significantly enhanced compared to the Schwarzschild case.
	Subsequently, we construct static thin-shell solutions by matching an interior wormhole geometry to an exterior vacuum spacetime across a junction surface. In the spirit of minimizing the presence of exotic matter, we identify parameter domains in which the null and weak energy conditions are satisfied at the shell. We examine the surface stress-energy components in detail, determining regions where the tangential surface pressure is either positive or negative, interpreted, respectively, as the pressure or surface tension. Additionally, an expression describing the behavior of the radial pressure across the junction is derived.
	Finally, we determine key geometrical characteristics of the wormhole, including the throat radius and the junction interface radius, by imposing traversability conditions. Estimates for the traversal time and required velocity are also provided, further elucidating the physical viability of these configurations.
	}
\keyword{wormholes; thin shells; stability}

\begin{document}




\section{Introduction}

Interest in traversable wormholes, which are hypothetical structures allowing for shortcuts through spacetime, has seen a significant resurgence following the seminal work of Morris and Thorne~\cite{Morris:1988cz}. In their influential paper, they provided a rigorous theoretical framework for static and spherically symmetric wormhole geometries that, in principle, could be traversed by human beings without encountering singularities or event horizons. This groundbreaking study not only revitalized discussions surrounding exotic spacetime configurations within general relativity but also laid the foundation for a broad array of theoretical investigations.
One of the most critical implications of traversable wormhole models is their unavoidable connection with violations of the classical energy conditions. In particular, the null energy condition (NEC) must be violated in order to sustain the throat of a traversable wormhole, implying the necessity of so-called ``exotic matter''. This has led to an extensive analyses of the energy condition violations and their physical plausibility, as reviewed in Refs.~\cite{Visser:1995cc,Lobo:2017cay,Barcelo:2002bv}, where the authors examined various quantum field theoretical and semiclassical sources that might support such exotic structures.

Additionally, the study of traversable wormholes has inspired discussions regarding closed timelike curves (CTCs)~\cite{Morris:1988tu,Krasnikov:2001cj} (for instance, see Refs.~\cite{Visser:1995cc,Lobo:2017cay} and references therein), raising profound questions about the nature of causality in spacetimes that admit such features. The potential emergence of CTCs in certain wormhole configurations highlights the tension between general relativity and the causal structure of spacetime, as explored in detailed analyses in Ref.~\cite{Visser:1995cc}.
Moreover, these investigations have extended into the realm of faster-than-light travel~\cite{Krasnikov:1995ad,Alcubierre:1994tu}. Theoretical constructions suggest that certain wormhole solutions might allow effective superluminal transit between distant regions of spacetime, at least from the perspective of external observers. This line of inquiry has been developed further in Ref.~\cite{Olum:1998mu}, where the author examined the physical characteristics and feasibility of such superluminal connections, emphasizing both their conceptual appeal and the formidable theoretical challenges they present.

The violation of the classical energy conditions remains a contentious and problematic aspect of traversable wormhole physics, its acceptability often contingent on one's theoretical perspective. Consequently, it is of considerable interest to explore configurations that minimize the use of exotic matter, which is defined as matter that violates the null energy condition. An elegant and insightful approach to this issue involves the construction of a class of wormhole solutions utilizing the ``cut-and-paste'' technique, introduced by \mbox{Visser~\cite{Visser:1995cc,Visser:1989kh,Visser:1989kg}}. This method allows for the concentration of exotic matter entirely at the wormhole throat, thereby localizing its effects and simplifying the overall spacetime geometry.
In such constructions, the surface stress--energy tensor components at the junction surface—where the exotic matter resides—are determined through the application of the Darmois--Israel formalism~\cite{Darmois,Israel:1966rt}. The resulting configurations, known as thin-shell wormholes, have proven particularly valuable as they lend themselves naturally to stability analyses under dynamic perturbations.

In fact, stability issues can be examined either by postulating specific surface equations of state~\cite{Kim:1992sh,Visser:1989am,Kim:1993bz} or more generally through a linearized stability analysis around a static equilibrium configuration~\cite{Poisson:1995sv,Lobo:2003xd}. In the latter approach, a parametrization of the stability conditions is employed, allowing one to bypass the need for a specified equation of state at the throat.
An extension of this formalism was presented in~\cite{Lobo:2003xd}, where the Poisson--Visser linearized stability framework~\cite{Poisson:1995sv} was generalized to include the presence of a non-zero cosmological constant. In particular, it was shown that for large positive values of the cosmological constant—corresponding to the Schwarzschild–de Sitter geometry—the domain of stability for thin-shell wormholes is significantly enhanced compared to the Schwarzschild case originally studied by Poisson and Visser. This result suggests that a positive cosmological constant can play a stabilizing role in the dynamics of thin-shell wormhole configurations.

As an alternative to thin-shell wormhole constructions, one may consider models in which the exotic matter is not confined to an infinitesimal shell but rather distributed throughout a finite region extending from the wormhole throat to a finite radius $a$. Beyond this radius, the interior solution is matched smoothly to an exterior vacuum spacetime. While several illustrative cases were initially discussed by Morris and Thorne~\cite{Morris:1988cz}, the analysis can be significantly broadened by employing the Darmois--Israel formalism~\cite{Israel:1966rt}.
In these models, the exotic matter threading the wormhole is localized within the interior region $r< a$, while the thin shell at the junction surface contributes a delta-function distribution of stress energy. This approach provides a natural framework for minimizing the amount of exotic matter required to support the wormhole geometry. Indeed, it was shown that, in principle, the quantity of exotic matter can be made arbitrarily small~\cite{Visser:2003yf}, provided that appropriate matching conditions are satisfied. Furthermore, one may impose that the surface stress-energy components of the thin shell obey the standard energy conditions, thereby enhancing the physical plausibility of the configuration.

A systematic generalization of such composite wormhole models was undertaken in~\cite{Lobo:2004id,Lobo:2004rp}, where the analysis extended the particular scenario studied in Ref.~\cite{Lemos:2003jb}, in which the redshift function is constant and the surface energy density at the junction surface vanishes. This framework allows for the classification of broader families of wormhole solutions based on specific interior geometries and matching conditions.
A similar construction has been applied to planar symmetric wormholes in the presence of a negative cosmological constant~\cite{Lemos:2004vs,Lemos:2008aj}. These planar symmetric wormholes represent a natural generalization of the topological black hole solutions originally discovered in Refs.~\cite{Lemos:1994fn,Lemos:1995cm,Lemos:1997bd}, wherein the introduction of exotic matter modifies the black hole horizon structure into a traversable wormhole geometry. Such configurations may possess planar topology or, upon compactification of one or more spatial coordinates, exhibit cylindrical or toroidal topology. In this context, planar symmetric wormholes can be interpreted as domain walls or thin membranes connecting different asymptotic regions or even distinct universes, offering intriguing possibilities in the study of spacetime topology and gravitational physics.

\textls[-15]{In fact, thin-shell wormholes have emerged as a rich field of research within gravitational physics, offering a simplified yet powerful framework to study the geometry and stability of traversable wormholes. In addition to the works mentioned above, foundational studies have explored general dynamical frameworks for spherically symmetric thin-shell wormholes in general relativity~\cite{Garcia:2011aa} and have developed stability analyses under radial perturbations~\cite{Lobo:2005zu, Eiroa:2008ky}. Diverse extensions have examined their behavior in alternative geometries and gravitational theories, including Einstein--Maxwell theory with a Gauss-Bonnet term~\cite{Thibeault:2005ha}, higher-dimensional spacetimes~\cite{Dias:2010uh, Rahaman:2006vg}, Brans--Dicke gravity~\cite{Eiroa:2008hv, Yue:2011cq}, $f(R)$ gravity~\cite{Eiroa:2015hrt, Eiroa:2016zjx,Rosa:2023olc}, and in hybrid metric--Palatini gravity~\cite{Rosa:2018jwp,Rosa:2021mln,Rosa:2021yym,Godani:2023jyx}, $f(R,T)$ gravity, and \mbox{extensions~\cite{Rosa:2021teg,Rosa:2022osy,Rosa:2023guo,Ganiyeva:2025qwz}}. Explorations into the role of different matter sources such as Chaplygin \mbox{gas~\cite{Eiroa:2007qz, Eiroa:2009hm, Bejarano:2011yz}}, generalized equations of state~\cite{Halilsoy:2013iza}, and dilaton or axion fields~\cite{Usmani:2010cd, Bejarano:2006uj} have revealed significant insights into the required energy conditions and stability regimes. Geometric generalizations include cylindrical~\cite{Eiroa:2004at, Eiroa:2009nn, Mazharimousavi:2014gpa}, asymmetric~\cite{Forghani:2018gza}, and global cosmic string configurations~\cite{Bejarano:2006uj}. Notably, stability considerations have been extended to thin shells associated with regular black holes~\cite{Halilsoy:2013iza}, and even black bounce scenarios~\cite{Lobo:2020kxn}. Moreover, the incorporation of higher-curvature corrections through Lovelock \mbox{gravity~\cite{Mehdizadeh:2015dta}}, Einstein--Yang--Mills-Gauss--Bonnet~\cite{Mazharimousavi:2010bf}, and Einstein--Cartan theory~\cite{Mehdizadeh:2018smu} has broadened the landscape of viable thin-shell wormhole solutions. These works collectively underscore the versatility and adaptability of the thin-shell formalism as a probe into both classical and modified gravity frameworks, providing critical insights into the interplay between geometry, matter content, and the fundamental constraints governing traversable wormholes.}

This paper is structured as follows. In Section \ref{sec2}, we provide a brief overview of the Darmois--Israel thin-shell formalism, which serves as the foundation for the analysis of junction conditions between distinct spacetime regions. In Section \ref{sec3}, we conduct a linearized stability analysis of thin-shell wormholes in the presence of a general cosmological constant, extending previous work on this subject. Section \ref{sec4} is devoted to the construction of a composite spacetime in which an interior wormhole solution is matched to an exterior vacuum geometry. Within this framework, we identify specific parameter regions where the surface stresses at the junction satisfy the classical energy conditions. We further investigate the physical features of the surface stress--energy tensor, focusing on the sign of the tangential surface pressure—which may correspond to either positive pressure or negative pressure (i.e., surface tension). Additionally, we derive an expression governing the discontinuity in the radial pressure across the junction surface. Lastly, we explore the traversability conditions of the wormhole, estimating the proper traversal time and the corresponding velocity required to cross the wormhole geometry. Our conclusions and final remarks are presented in Section \ref{secconclusion}.

\section{Overview of the Darmois--Israel Formalism}\label{sec2}

Consider two distinct spacetime manifolds, ${\cal M_+}$ and ${\cal
	M_-}$, with metrics given by $g_{\mu \nu}^+(x^{\mu}_+)$ and
$g_{\mu \nu}^-(x^{\mu}_-)$, in terms of independently defined
coordinate systems $x^{\mu}_+$ and $x^{\mu}_-$. The manifolds are
bounded by hypersurfaces $\Sigma_+$ and $\Sigma_-$, respectively,
with induced metrics $g_{ij}^+$ and $g_{ij}^-$. The hypersurfaces
are isometric, i.e., $g_{ij}^+(\xi)=g_{ij}^-(\xi)=g_{ij}(\xi)$, in
terms of the intrinsic coordinates, invariant under the isometry.
A single manifold ${\cal M}$ is obtained by gluing together ${\cal
	M_+}$ and ${\cal M_-}$ at their boundaries, i.e., ${\cal M}={\cal
	M_+}\cup {\cal M_-}$, with the natural identification of the
boundaries $\Sigma=\Sigma_+=\Sigma_-$.

The three holonomic basis vectors $e_{(i)}=\partial /\partial
\xi^i$ tangent to $\Sigma$ have the  components
$e^{\mu}_{(i)}|_{\pm}=\partial x_{\pm}^{\mu}/\partial \xi^i$,
which provide the induced metric on the junction surface by the
following scalar product
\begin{equation}
	g_{ij}=e_{(i)}\cdot e_{(j)}=g_{\mu
		\nu}e^{\mu}_{(i)}e^{\nu}_{(j)}|_{\pm}.
\end{equation}

We shall consider a timelike junction surface $\Sigma$, defined by
the parametric equation of the form $f(x^{\mu}(\xi^i))=0$. The
unit normal $4-$vector, $n^{\mu}$, to $\Sigma$ is defined as
\begin{equation}\label{defnormal}
	n_{\mu}=\pm \,\left |g^{\alpha \beta}\,\frac{\partial f}{\partial
		x ^{\alpha}} \, \frac{\partial f}{\partial x ^{\beta}}\right
	|^{-1/2}\;\frac{\partial f}{\partial x^{\mu}},
\end{equation}
with $n_{\mu}\,n^{\mu}=+1$ and $n_{\mu}e^{\mu}_{(i)}=0$. The
Israel formalism requires that the normals point from ${\cal M_-}$
to ${\cal M_+}$~\cite{Israel:1966rt}.

\textls[-25]{The extrinsic curvature, or the second fundamental form, is
defined as $K_{ij}=n_{\mu;\nu}e^{\mu}_{(i)}e^{\nu}_{(j)}$, or}
\begin{eqnarray}\label{extrinsiccurv}
	K_{ij}^{\pm}=-n_{\mu} \left(\frac{\partial ^2 x^{\mu}}{\partial
		\xi ^{i}\,\partial \xi ^{j}}+\Gamma ^{\mu \pm}_{\;\;\alpha
		\beta}\;\frac{\partial x^{\alpha}}{\partial \xi ^{i}} \,
	\frac{\partial x^{\beta}}{\partial \xi ^{j}} \right).
\end{eqnarray}
Note 
 that for the case of a thin shell, $K_{ij}$ is not continuous
across $\Sigma$, so that for notational convenience, the
discontinuity in the second fundamental form is defined as
$\kappa_{ij}=K_{ij}^{+}-K_{ij}^{-}$.

Now, the Lanczos equations follow from the Einstein equations for
the hypersurface  and are given by
\begin{equation}
	S^{i}_{\;j}=-\frac{1}{8\pi}\,(\kappa ^{i}_{\;j}-\delta
	^{i}_{\;j}\kappa ^{k}_{\;k}),
\end{equation}
where $S^{i}_{\;j}$ is the surface stress--energy tensor on
$\Sigma$.

The first contracted Gauss--Kodazzi equation, or the ``Hamiltonian''
constraint,
\begin{eqnarray}
	G_{\mu \nu}n^{\mu}n^{\nu}=\frac{1}{2}\,(K^2-K_{ij}K^{ij}-\,^3R),
\end{eqnarray}
and the Einstein equations provide the evolution identity
\begin{eqnarray}
	S^{ij}\overline{K}_{ij}=-\left[T_{\mu
		\nu}n^{\mu}n^{\nu}-\Lambda/8\pi \right].
\end{eqnarray}
The convention $\left[X \right]\equiv X^+|_{\Sigma}-X^-|_{\Sigma}$
and $\overline{X} \equiv (X^+|_{\Sigma}+X^-|_{\Sigma})/2$ is used.

The second contracted Gauss--Kodazzi equation, or the ``ADM''
constraint,
\begin{eqnarray}
	G_{\mu \nu}e^{\mu}_{(i)}n^{\nu}=K^j_{i|j}-K,_{i},
\end{eqnarray}
with the Lanczos equations gives the conservation identity
\begin{eqnarray}
	S^{i}_{j|i}=\left[T_{\mu \nu}e^{\mu}_{(j)}n^{\nu}\right].
\end{eqnarray}

In particular, considering spherical symmetry, considerable
simplifications occur, namely $\kappa ^{i}_{\;j}={\rm diag}
\left(\kappa ^{\tau}_{\;\tau},\kappa ^{\theta}_{\;\theta},\kappa
^{\theta}_{\;\theta}\right)$. The surface stress--energy tensor may
be written in terms of the surface energy density, $\sigma$, and
the surface pressure, $p$, as $S^{i}_{\;j}={\rm
	diag}(-\sigma,p,p)$. The Lanczos equations then reduce to
\begin{eqnarray}
	\sigma &=&-\frac{1}{4\pi}\,\kappa ^{\theta}_{\;\theta} ,\label{sigma} \\
	p &=&\frac{1}{8\pi}(\kappa ^{\tau}_{\;\tau}+\kappa
	^{\theta}_{\;\theta}). \label{surfacepressure}
\end{eqnarray}

\section{Cut-and-Paste Technique: Thin-Shell Wormholes with a \mbox{Cosmological Constant}} \label{sec3}

In this section we shall construct a class of wormhole solutions,
in the presence of a cosmological constant, using the
cut-and-paste technique. Consider the unique spherically symmetric
vacuum solution, i.e.,
\begin{eqnarray}
	ds^2=-\left(1-\frac{2M}{r}-\frac{\Lambda}{3}r^2
	\right)\,dt^2+\left(1-\frac{2M}{r}-\frac{\Lambda}{3}r^2
	\right)^{-1}\,dr^2 +r^2\,(d\theta ^2+\sin ^2{\theta}\, d\phi ^2)
	\label{metricvacuumlambda}.
\end{eqnarray}
If $\Lambda >0$, the solution is denoted by the Schwarzschild--de
Sitter metric. For $\Lambda <0$, we have the Schwarzschild--anti-de
Sitter metric, and of course the specific case of $\Lambda =0$ is
reduced to the Schwarzschild solution. Note that the metric $(1)$
is not asymptotically flat as $r \rightarrow \infty$. Rather, it
is asymptotically de Sitter if $\Lambda >0$, or asymptotically
anti-de Sitter, if $\Lambda <0$. But, considering low values of
$\Lambda$, the metric is almost flat in the range $M \ll r \ll
1/\sqrt{\Lambda}$. For values below this range, the effects of $M$
dominate, and for values above the range, the effects of $\Lambda$
dominate, as for large values of the radial coordinate the
large-scale structure of the spacetime must be taken into account.

The specific case of $\Lambda =0$ is reduced to the Schwarzschild
solution, with a black hole event horizon at $r_b=2M$. Consider
the Schwarzschild--de Sitter spacetime, $\Lambda
>0$. If $0<9\Lambda M^2<1$, the factor $g(r)=(1-2M/r-\Lambda r^2/3)$
possesses two positive real roots, $r_b$ and $r_c$, corresponding
to the black hole and the cosmological event horizons of the de
Sitter spacetime, respectively, given by
\begin{eqnarray}
	r_b&=&2 \Lambda ^{-1/2} \, \cos(\alpha/3)    \label{root1}  , \\
	r_c&=&2 \Lambda ^{-1/2} \, \cos(\alpha/3+4\pi/3)    \label{root2}
	,
\end{eqnarray}
where $\cos \alpha \equiv -3M \Lambda^{1/2}$, with $\pi < \alpha
<3\pi/2$. In this domain, we have $2M<r_b<3M$ and $r_c>3M$.

\textls[+25]{For the Schwarzschild--anti-de Sitter metric, with $\Lambda <0$,
the factor $g(r)=\left(1-2M/r+|\Lambda |r^2/3 \right)$} has only
one real positive root, $r_b$, given by
\begin{eqnarray}
	r_b=\left(\frac{3M}{|\Lambda|}\right)^{1/3}\left(\sqrt[3]{1+\sqrt{1+\frac{1}{9|\Lambda|M^2}}}
	+\sqrt[3]{1-\sqrt{1+ \frac{1}{9|\Lambda|M^2}}}\;\right)
	\label{adsbhole},
\end{eqnarray}
corresponding to a black hole event horizon, with $0<r_b<2M$.

\subsection{The Cut-and-Paste Construction}

Given this, we may construct a wormhole solution using the
cut-and-paste technique~\cite{Visser:1995cc,Visser:1989kh,Visser:1989kg}.
Consider two vacuum solutions with $\Lambda$ and remove from each
spacetime the region described by
\begin{equation}
	\Omega_{\pm}\equiv \left \{r_{\pm}\leq a| \,a >r_b \right \} ,
\end{equation}
where $a$ is a constant and $r_b$ is the black hole event horizon,
corresponding to the Schwarzschild--de Sitter and
Schwarzschild--anti-de Sitter solutions, Equation (\ref{root1}) and
Equation (\ref{adsbhole}), respectively. The removal of the
regions results in two geodesically incomplete manifolds, with
boundaries given by the following timelike hypersurfaces
\begin{equation}
	\Sigma_{\pm}\equiv \left \{r_{\pm}= a| \,a > r_b \right \} .
\end{equation}
Identifying these two timelike hypersurfaces,
$\Sigma_{+}=\Sigma_{-}$, results in a geodesically complete
manifold, with two regions connected by a wormhole and the
respective throat situated at $\Sigma$. The wormhole connects two
regions, asymptotically de Sitter or anti-de Sitter, for $\Lambda
>0$ and $\Lambda <0$, respectively.

The intrinsic metric at $\Sigma$ is given by
\begin{equation}
	ds^2_{\Sigma}=-d\tau ^2 +a^2(\tau)\,(d\theta ^2+\sin ^2{\theta}\,
	d\phi ^2),
\end{equation}
where $\tau$ is the proper time as measured by a comoving observer
on the wormhole throat.

\subsection{The Surface Stresses}

The imposition of spherical symmetry is sufficient to conclude
that there is no gravitational radiation, independently of the
behavior of the wormhole throat. The position of the throat is
given by
$x^{\mu}(\tau,\theta,\phi)=(t(\tau),a(\tau),\theta,\phi)$, and the
respective $4$-velocity is
\begin{equation}
	U^{\mu}=\left(\frac{\sqrt{1-2M/a-\Lambda a^2/3+\dot{a}^2}}
	{1-2M/a-\Lambda a^2/3},\dot{a},0,0 \right)  ,
\end{equation}
where the overdot denotes a derivative with respect to $\tau$.

The unit normal to the throat may be determined by Equation
(\ref{defnormal}) or by the contractions, $U^{\mu}n_{\mu}=0$ and
$n^{\mu}n_{\mu}=+1$ and is given by
\begin{equation}
	n^{\mu}=\left(\frac{\dot{a}} {1-2M/a-\Lambda
		a^2/3},\sqrt{1-2M/a-\Lambda a^2/3+\dot{a}^2},0,0 \right)
	\label{normal} .
\end{equation}

Using Equations (\ref{extrinsiccurv}), 
(\ref{metricvacuumlambda}) and  (\ref{normal}), the
non-trivial components of the extrinsic curvature are given by
\begin{eqnarray}
	K^{\theta\;\pm}_{\;\,\theta}&=& \pm
	\frac{1}{a}\sqrt{1-2M/a-\Lambda a^2/3+\dot{a}^2} ,
	\label{Kplustheta}\\
	K ^{\tau \;\pm}_{\;\,\tau}&=& \pm \frac{M/a^2-\Lambda a/3+
		\ddot{a}} {\sqrt{1-2M/a-\Lambda a^2/3+\dot{a}^2}} ,
	\label{Kplustautau}
\end{eqnarray}
Thus, the Einstein field equations, Equations
(\ref{sigma}) and (\ref{surfacepressure}), with Equations
(\ref{Kplustheta}) and (\ref{Kplustautau}), then provide us with the
following surface stresses:
\begin{eqnarray}
	\sigma&=&-\frac{1}{2\pi a} \sqrt{1-2M/a-\Lambda a^2/3+\dot{a}^2}
	\label{surfenergy1}   ,\\
	p&=&\frac{1}{4\pi a} \;\frac{1-M/a-2\Lambda
		a^2/3+\dot{a}^2+a\ddot{a}}{\sqrt{1-2M/a-\Lambda a^2/3+\dot{a}^2}}
	\label{surfpressure1}  .
\end{eqnarray}

We also verify that the above equations imply the conservation of
the surface stress--energy tensor
\begin{equation}
	\dot{\sigma}=-2 \left(\sigma +p \right) \frac{\dot{a}}{a}
	\label{conservationenergy}
\end{equation}
or
\begin{equation}
	\frac{d\left(\sigma A \right)}{d\tau}+p\,\frac{dA}{d\tau}=0 ,
\end{equation}
where $A=4\pi a^2$ is the area of the wormhole throat. The first
term represents the variation of the internal energy of the
throat, and the second term is the work performed out by the throat's
internal forces.

\subsection{Linearized Stability Analysis}

Equation (\ref{surfenergy1}) may be recast into the following
dynamical form
\begin{equation}
	\dot{a}^2-\frac{2M}{a}-\frac{\Lambda}{3} a^2- \left(2\pi \sigma a
	\right)^2=-1  ,  \label{motionofthroat}
\end{equation}
which determines the motion of the wormhole throat. Considering an
equation of state of the form, $p=p(\sigma)$, the energy
conservation, Equation (\ref{conservationenergy}), can be
integrated to yield
\begin{equation}
	\ln(a)=-\frac{1}{2}\int \frac{d\sigma}{\sigma +p(\sigma)}  \,.
\end{equation}
\textls[-25]{This result, formally inverted to provide $\sigma=\sigma(a)$, can
be finally substituted into \mbox{Equation (\ref{motionofthroat})}}. The
latter can also be written as $\dot{a}^2=-V(a)$, with the
potential defined as
\begin{equation}
	V(a)=1-\frac{2M}{a}-\frac{\Lambda}{3} a^2- \left(2\pi \sigma a
	\right)^2     \label{defpotential}  .
\end{equation}

One may explore specific equations of state, but following the
Poisson--Visser \mbox{analysis~\cite{Poisson:1995sv}}, we shall consider a linear
perturbation around a static solution with radius $a_0$. The
respective values of the surface energy density and the surface
pressure, at $a_0$, are given by
\begin{eqnarray}
	\sigma_0&=&-\frac{1}{2\pi a_0} \sqrt{1-2M/a_0-\Lambda
		a_0^2/3}   \label{stablesigma} \,,\\
	p_0&=&\frac{1}{4\pi a_0}\;\frac{1-M/a_0-2\Lambda
		a_0^2/3}{\sqrt{1-2M/a_0-\Lambda a_0^2/3}} \,.
	\label{stablepressure}
\end{eqnarray}
One verifies that the surface energy density is always negative,
implying the violation of the weak and dominant energy conditions.
One may verify that the null and strong energy conditions are
satisfied for $a_0 \leq 3M$ for a generic $\Lambda$, for a generic
cosmological \mbox{constant~\cite{Lobo:2003xd}}.

Linearizing around the stable solution at $a=a_0$, we consider a
Taylor expansion of $V(a)$ around $a_0$ to the second order, which
provides
\begin{equation}
	V(a)=V(a_0)+V'(a_0)(a-a_0)+\frac{1}{2}\,V''(a_0)(a-a_0)^2+O
	\left[(a-a_0)^3 \right] \label{Taylorexpansion} ,
\end{equation}
where the prime denotes a derivative with respect to $a$, $d/da$.

Define the parameter $\eta_s(\sigma)=dp/d\sigma=p'/\sigma'$. The
physical interpretation of $\eta_s$ is discussed in~\cite{Poisson:1995sv},
and $\sqrt{\eta_s}$ is normally interpreted as the speed of sound.
Evaluated at the static solution, at $a=a_0$, using Equations
(\ref{stablesigma}) and (\ref{stablepressure}), we readily find
$V(a_0)=0$ and $V'(a_0)=0$, with $V''(a_0)$ given by
\begin{equation}
	V''(a_0)=-\frac{2}{a_0^2} \Bigg[
	\frac{2M}{a_0}+\frac{\Lambda}{3}a_0^2+\frac{\left(M/a_0-\Lambda
		a_0^2/3 \right)^2}{1-2M/a_0-\Lambda a_0^2/3}+(1+2\eta)
	\left(1-\frac{3M}{a_0} \right) \Bigg],
\end{equation}
where $\eta=\eta_s(\sigma_0)$.

The potential $V(a)$, Equation (\ref{Taylorexpansion}), is reduced
to
\begin{equation}
	V(a)=\frac{1}{2}\,V''(a_0)(a-a_0)^2+O \left[(a-a_0)^3 \right],
\end{equation}
so that the equation of motion for the wormhole throat presents
the following form:
\begin{equation}
	\dot{a}^2=-\frac{1}{2}V''(a_0)(a-a_0)^2+O \left[(a-a_0)^3 \right]
	,
\end{equation}
to the order of approximation considered. If $V''(a_0)<0$ is
verified, then the potential $V(a_0)$ has a local maximum at
$a_0$, where a small perturbation in the wormhole throat's radius
will provoke an irreversible contraction or expansion of the
throat. Thus, the solution is stable if and only if $V(a_0)$ has a
local minimum at $a_0$ and $V''(a_0)>0$, i.e.,
\begin{equation}
	\eta \left(1-\frac{3M}{a_0} \right) <
	-\frac{1-3M/a_0+3M^2/a_0^2-\Lambda Ma_0}{2 \left(1-2M/a_0-\Lambda
		a_0^2/3 \right)}
	\label{inequality}.
\end{equation}

\textls[-15]{We need to analyze Equation (\ref{inequality}) for several cases.
The right-hand side of \mbox{Equation (\ref{inequality})}} is always
negative~\cite{Lobo:2003xd}, while the left-hand side changes sign
at $a_0=3M$. Thus, one deduces that the stability regions are
dictated by the following inequalities:
\begin{eqnarray}
	\eta <-\frac{1-3M/a_0+3M^2/a_0^2-\Lambda
		Ma_0}{2\left(1-2M/a_0-\Lambda a_0^2/3 \right)\left(1-3M/a_0
		\right)}\,,\qquad& \,a_0>3M         \label{eta01}
	\\
	\eta >-\frac{1-3M/a_0+3M^2/a_0^2-\Lambda
		Ma_0}{2\left(1-2M/a_0-\Lambda a_0^2/3 \right)\left(1-3M/a_0
		\right)}\,,\qquad& \,a_0<3M .      \label{eta02}
\end{eqnarray}

One may analyze several cases.

\subsubsection{De Sitter Spacetime}

For the de Sitter spacetime, with $M=0$ and $\Lambda >0$, Equation
(\ref{inequality}) reduces to
\begin{equation}
	\eta <-\frac{1}{2\left(1-\Lambda a_0^2/3 \right)} \,, \qquad
	{\rm for}\qquad 0<a_0<\sqrt{3/\Lambda} .
\end{equation}
The stability region is depicted in the left plot of Figure \ref{fig1}.

\subsubsection{Anti-de Sitter Spacetime}

\begin{figure}[H]
	\includegraphics[width=2.1in]{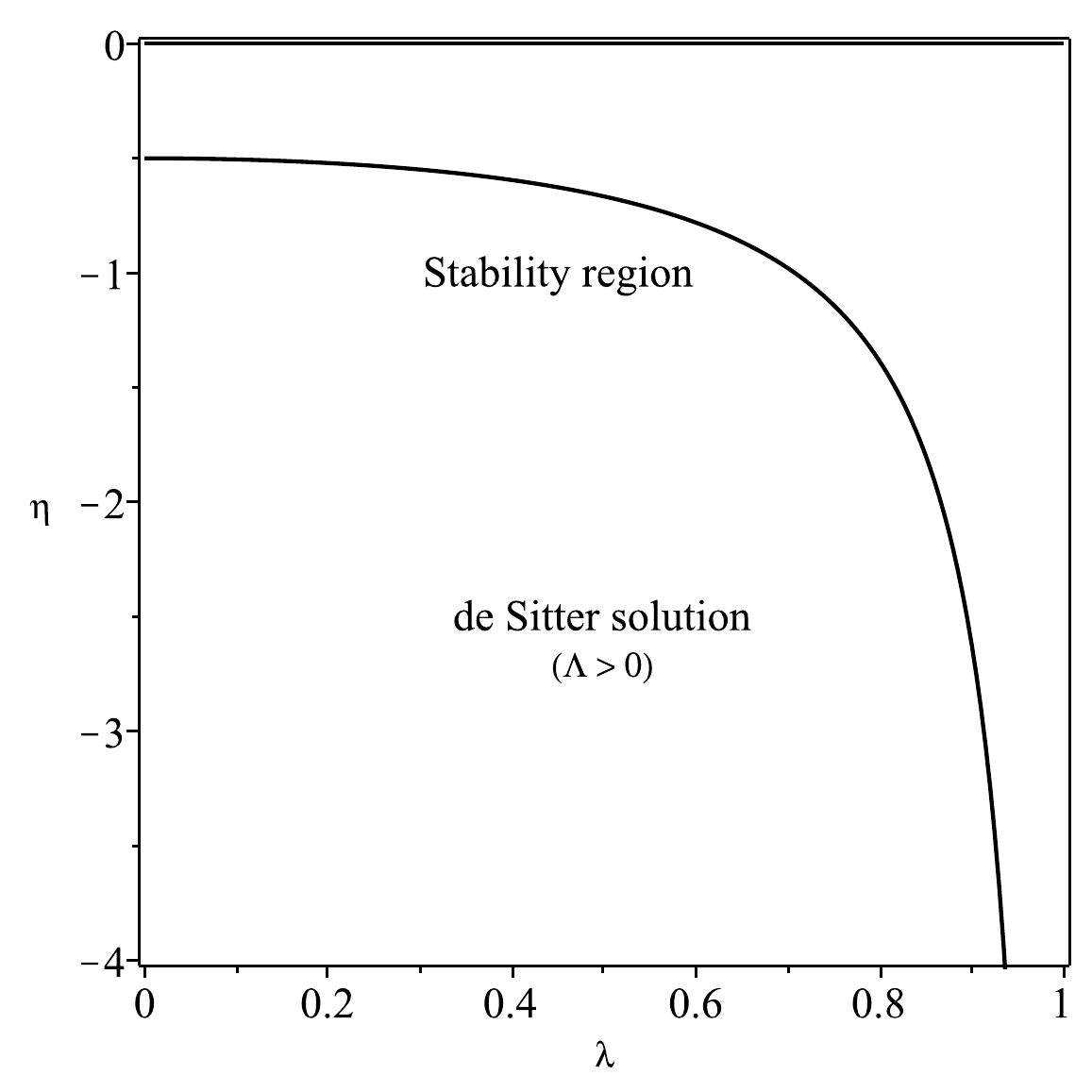}
	\hspace{0.4in}
	\includegraphics[width=2.1in]{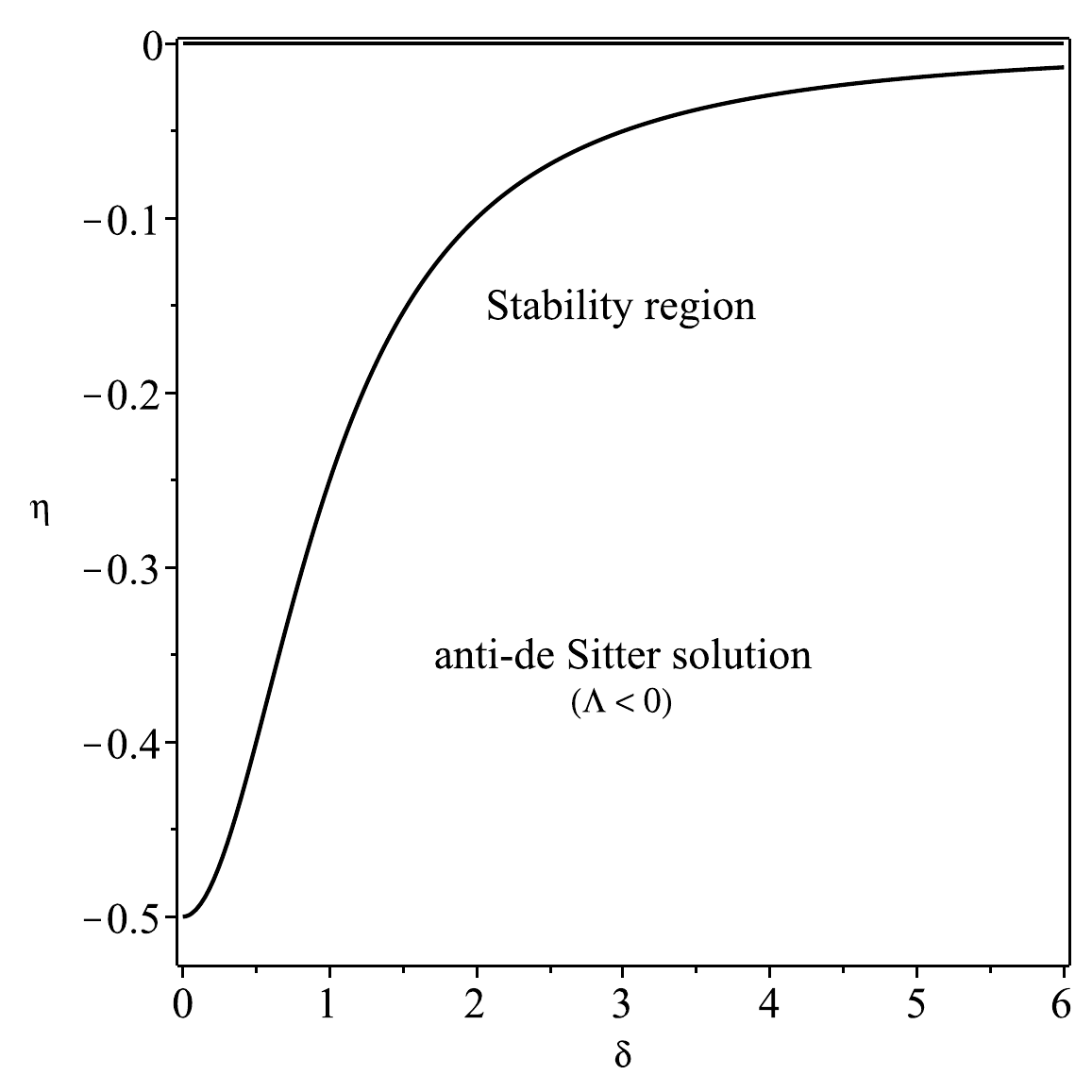}
	\caption{We 
 have defined $\lambda = a_0/(3/\Lambda)^{1/2}$ and
		$\delta = a_0/(3/|\Lambda|)^{1/2}$, respectively.
		The regions of stability are depicted in the graphs,
		below the curves for the de-Sitter and the anti-de Sitter \mbox{solutions, respectively} (plots adapted from Figure 3 of Ref.~\cite{Lobo:2003xd}).}
		\label{fig1}
\end{figure}

For the anti-de Sitter spacetime, with $M=0$ and $\Lambda <0$,
Equation (\ref{inequality}) gives
\begin{equation}
	\eta <-\frac{1}{2\left(1+|\Lambda| \,a_0^2/3 \right)} \,,
	\qquad {\rm for} \qquad  a_0>0   \label{adsconstraint},
\end{equation}
and the respective stability region is depicted in the right plot
of Figure \ref{fig1}.

\subsubsection{Schwarzschild Spacetime}

This is the particular case of the Poisson--Visser analysis
\cite{Poisson:1995sv}, with $\Lambda=0$, which reduces to
\begin{eqnarray}
	\eta <-\frac{1-3M/a_0+3M^2/a_0^2}{2\left(1-2M/a
		\right)\left(1-3M/a_0 \right)}\,, \qquad& a_0>3M ,  \\
	\eta >-\frac{1-3M/a_0+3M^2/a_0^2}{2\left(1-2M/a
		\right)\left(1-3M/a_0 \right)}\,, \qquad& a_0<3M .
\end{eqnarray}
The stability regions are shown in the left plot of Figure \ref{fig2}.

\subsubsection{Schwarzschild--Anti-de Sitter Spacetime}

For the Schwarzschild--anti-de Sitter spacetime, with $\Lambda <0$,
we have
\begin{eqnarray}
	\eta <-\frac{1-3M/a_0+3M^2/a_0^2+|\Lambda| \,
		Ma_0}{2\left(1-2M/a_0+|\Lambda|\, a_0^2/3 \right)\left(1-3M/a_0
		\right)}  \,, \qquad& a_0>3M     \label{ads1}
	\\
	\eta >-\frac{1-3M/a_0+3M^2/a_0^2+|\Lambda| \,
		Ma_0}{2\left(1-2M/a_0+|\Lambda|\, a_0^2/3 \right)\left(1-3M/a_0
		\right)}\,, \qquad& a_0<3M        \label{ads2}
\end{eqnarray}

\begin{figure}[H]
	\includegraphics[width=2in]{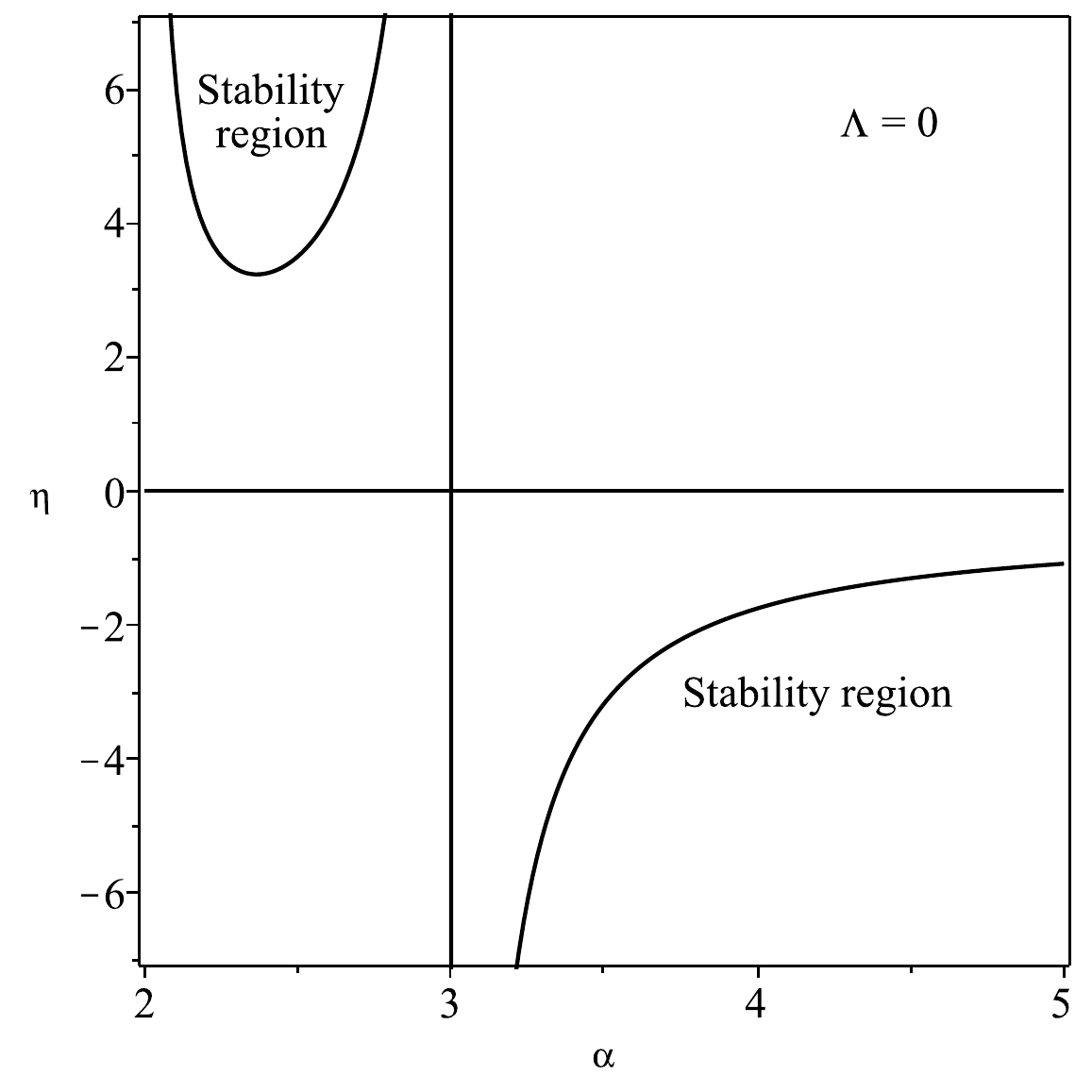}
	\hspace{0.4in}
	\includegraphics[width=2in]{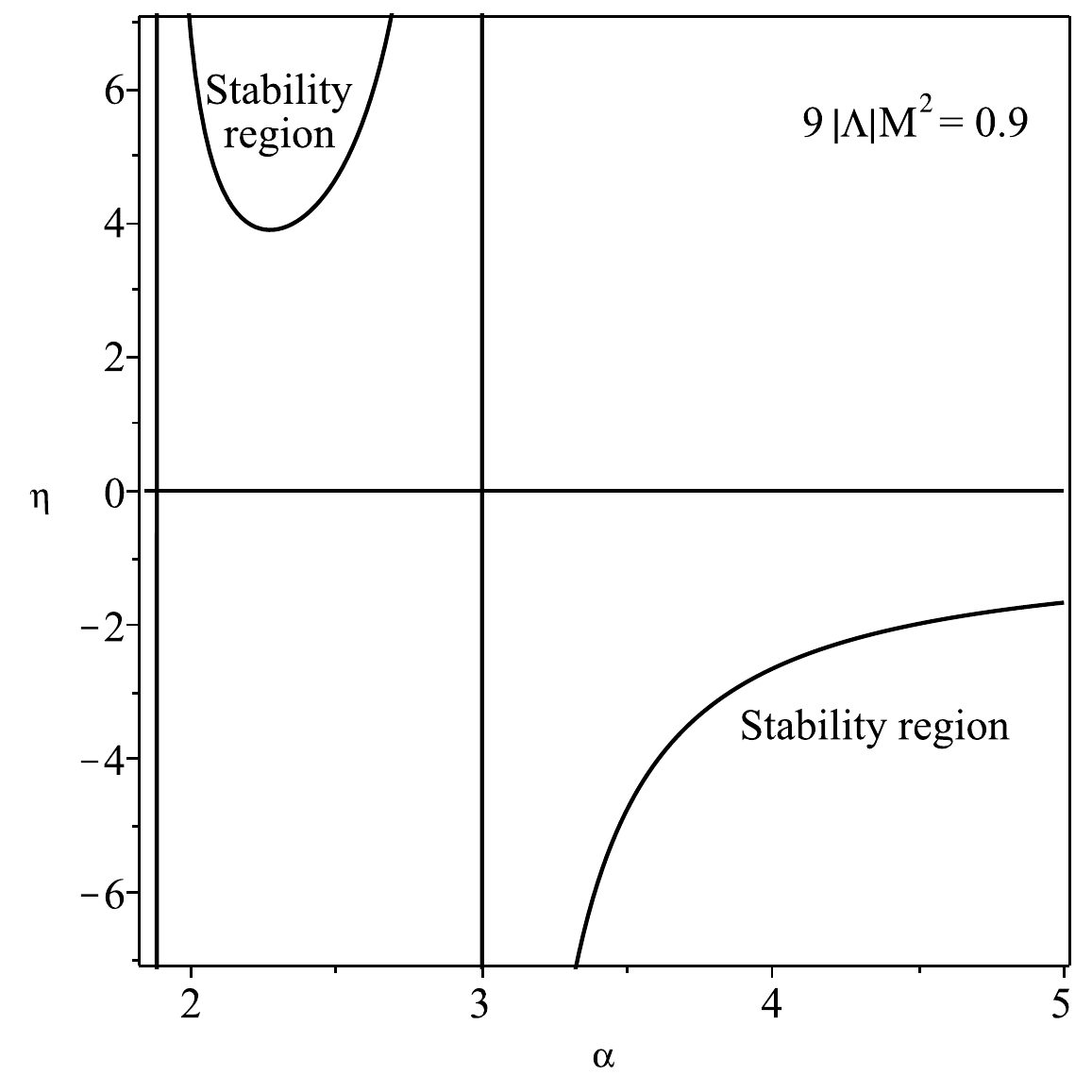}
	
	\vspace{-2pt}
	
	\caption{We have defined $\alpha =a_0/M$.
		The regions of stability are depicted for the
		Schwarzschild and the Schwarzschild--anti-de Sitter solutions, respectively.
		Imposing the value $9|\Lambda |M^2=0.9$ in the Schwarzschild--anti-de Sitter case,
		we verify that the stability regions decrease relative to the
		Schwarzschild solution (plots adapted from Figure 4 of Ref.~\cite{Lobo:2003xd}).}
		\label{fig2}
\end{figure}

The regions of stability are depicted in the right plot of Figure
\ref{fig2}, considering the value $9|\Lambda |M^2=0.9$. In this case, the
black hole event horizon is given by $r_b \simeq 1.8\,M$. We
verify that the regions of stability decrease relative to the
Schwarzschild case.

\vspace{-2pt}

\subsubsection{Schwarzschild--de Sitter Spacetime}

For the Schwarzschild--de Sitter spacetime, with $\Lambda
>0$, we have
\begin{eqnarray}
	\eta < -\frac{1-3M/a_0+3M^2/a_0^2-\Lambda
		Ma_0}{2\left(1-2M/a_0-\Lambda a_0^2/3 \right)\left(1-3M/a_0
		\right)}   \,,    \qquad&   \,a_0>3M
	\\
	\eta > -\frac{1-3M/a_0+3M^2/a_0^2-\Lambda
		Ma_0}{2\left(1-2M/a_0-\Lambda a_0^2/3 \right)\left(1-3M/a_0
		\right)}  \,,    \qquad&     \,a_0<3M .
\end{eqnarray}

\begin{figure}[H]
	\includegraphics[width=2.4in]{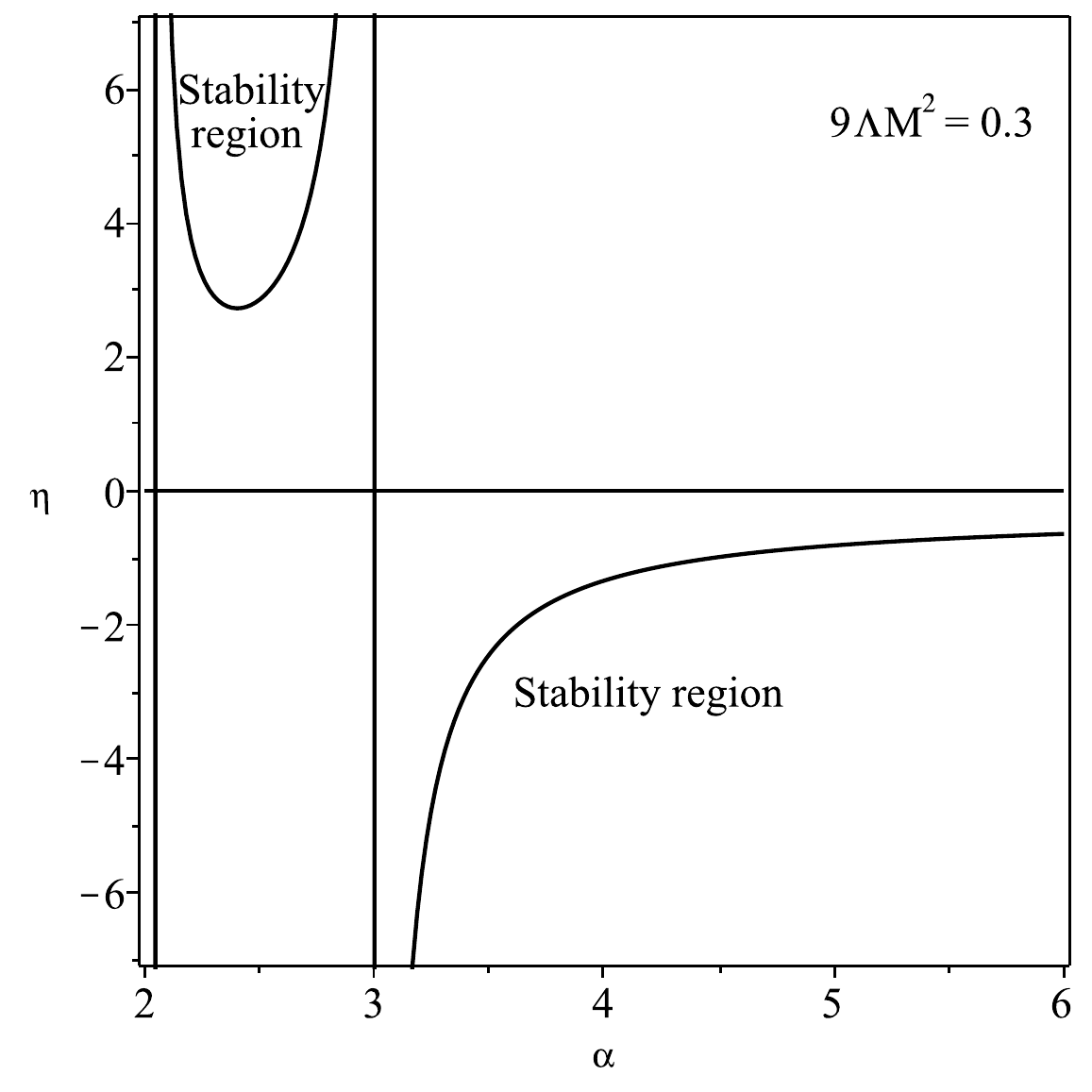}
	\hspace{0.4in}
	\includegraphics[width=2.4in]{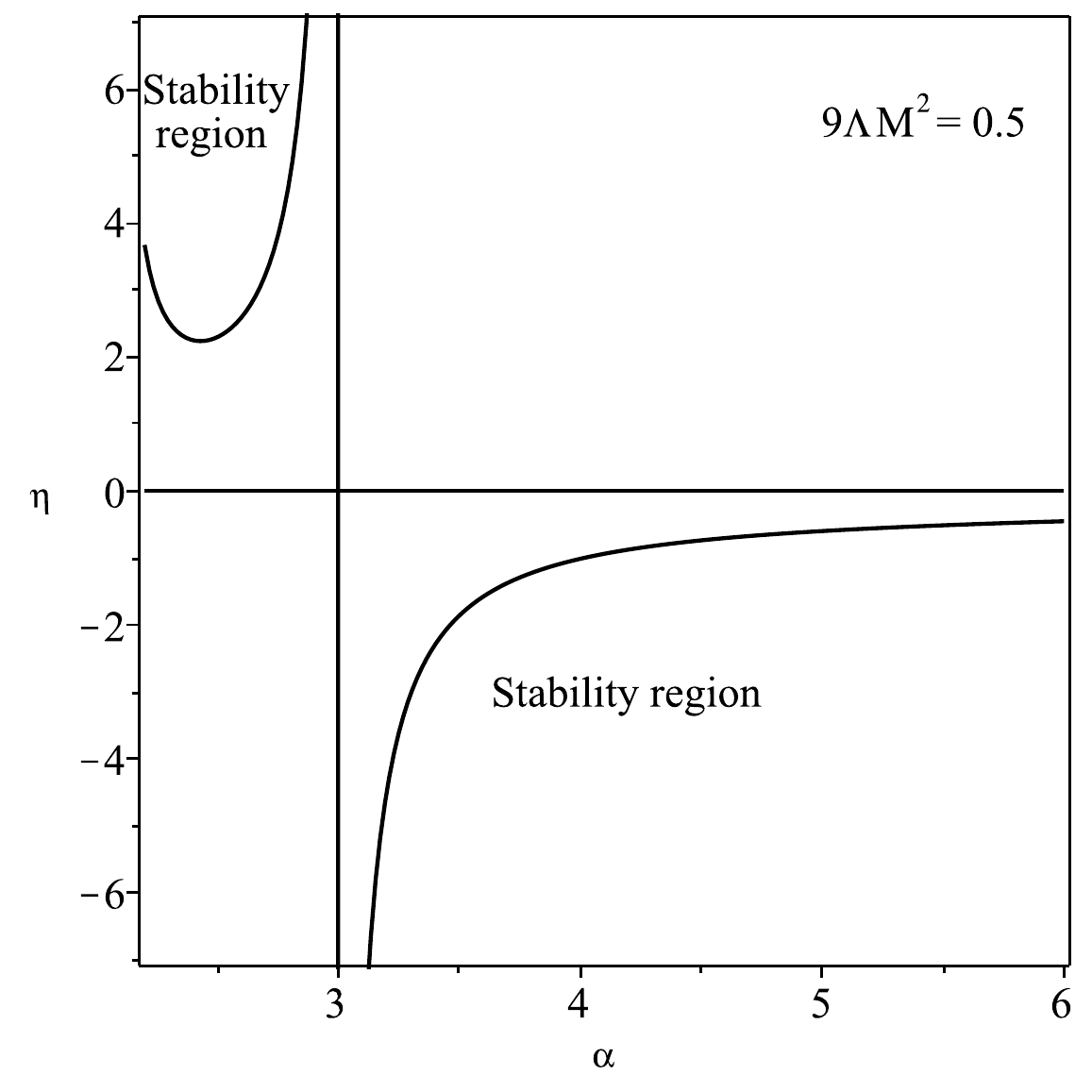}
%
	\includegraphics[width=2.4in]{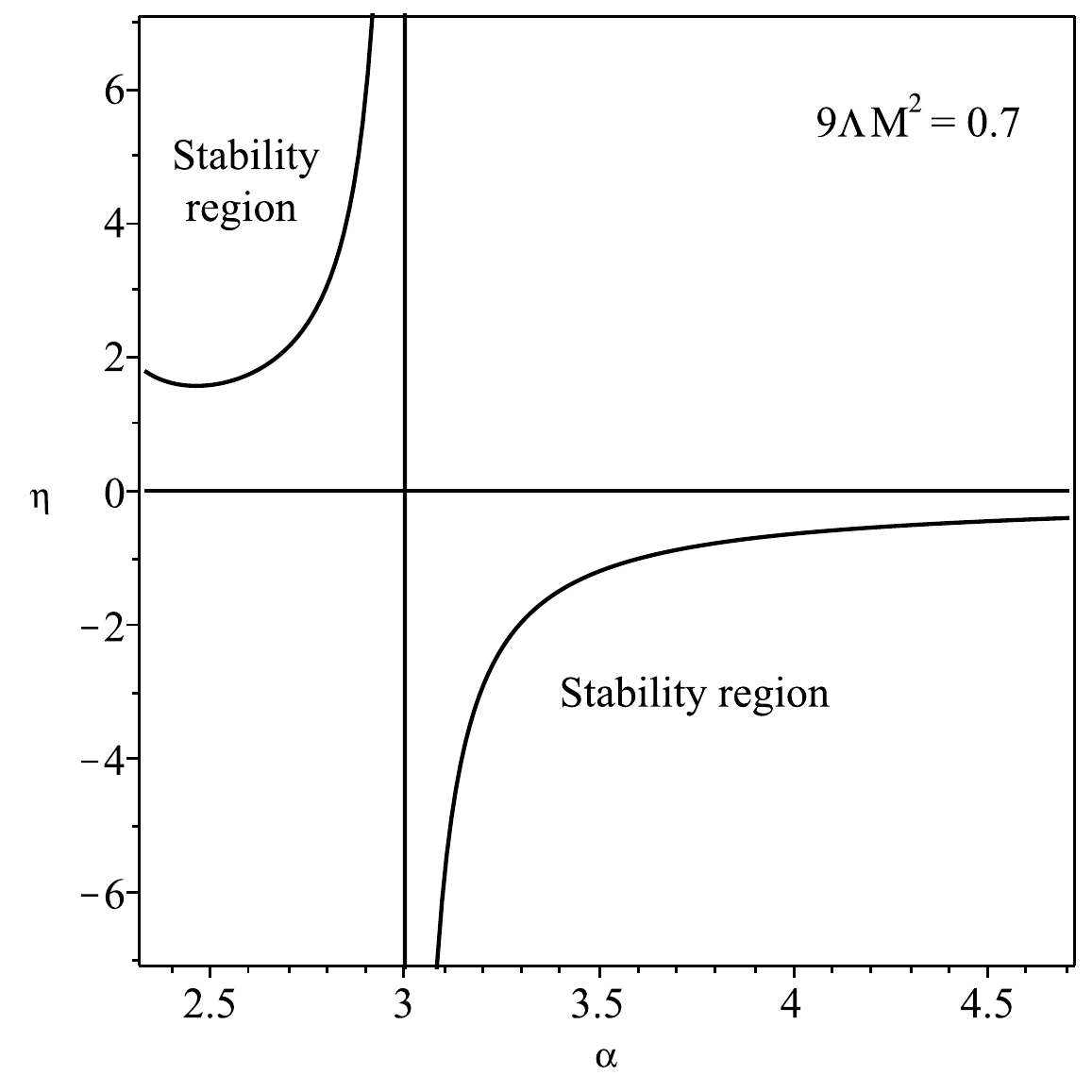}
	\hspace{0.4in}
	\includegraphics[width=2.4in]{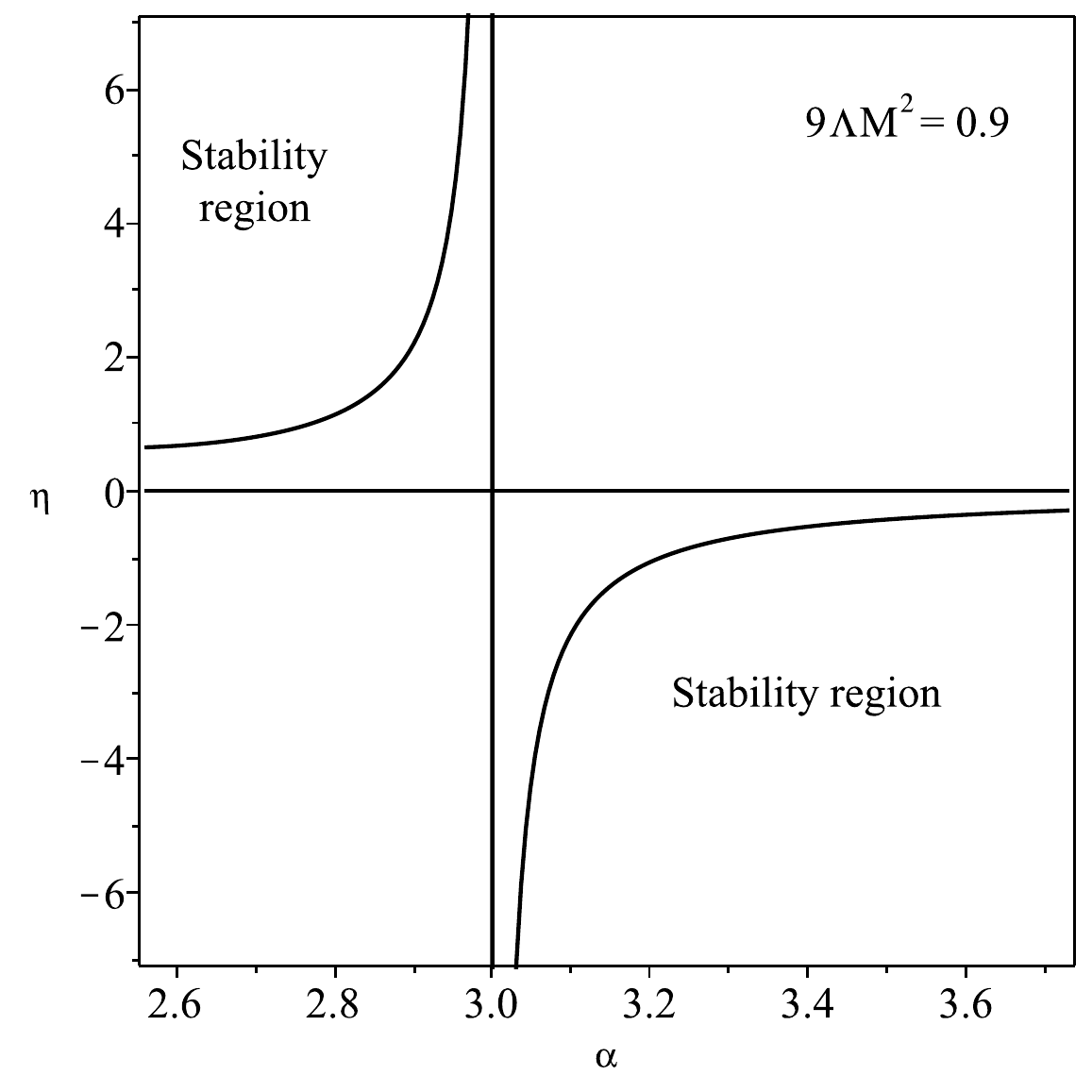}
	\caption{We have defined $\alpha =a_0/M$.
		The regions of stability for the Schwarzschild--de Sitter solution,
		imposing $9\Lambda M^2=0.3$, $9\Lambda M^2=0.5$, $9\Lambda M^2=0.7$
		and $9\Lambda M^2=0.9$, respectively.
		The regions of stability are significantly increased, relative
		to the $\Lambda =0$ case, for increasing values of $9\Lambda M^2$ (plots adapted from Figure 5 of Ref.~\cite{Lobo:2003xd}).}
		\label{fig3}
\end{figure}

The regions of stability are depicted in Figure \ref{fig3} for increasing
values of $9\Lambda M^2$. In particular, for $9\Lambda M^2=0.7$,
the black hole and cosmological horizons are given by \mbox{$r_b \simeq
2.33\,M$ and $r_c =4.71\,M$,} respectively. Thus, only the interval
$2.33 < a_0/M < 4.71$ is taken into account, as shown in the range
of the respective plot. Analogously, for $9\Lambda M^2=0.9$, we
find $r_b \simeq 2.56\,M$ and $r_c =3.73\,M$. Therefore, only the
range within the interval $2.56 < a_0/M < 3.73$ corresponds to the
stability regions, also shown in the respective plot.

We verify that for large values of $\Lambda$, or large $M$, the
regions of stability are significantly increased  relative to
the $\Lambda =0$ case.

\section{A Thin Shell Around a Traversable Wormhole}\label{sec4}

As an alternative to the thin-shell wormhole, constructed using
the cut-and-paste technique analyzed above, one may also consider
that the exotic matter is distributed from the throat $r_0$ to a
radius $a$, where the solution is matched to an exterior vacuum
spacetime. Thus, the thin shell confines the exotic matter
threading the wormhole to a finite region, with a delta function
distribution of the stress--energy tensor on the junction surface.

\begin{figure}[H]
	\includegraphics[width=3.5in]{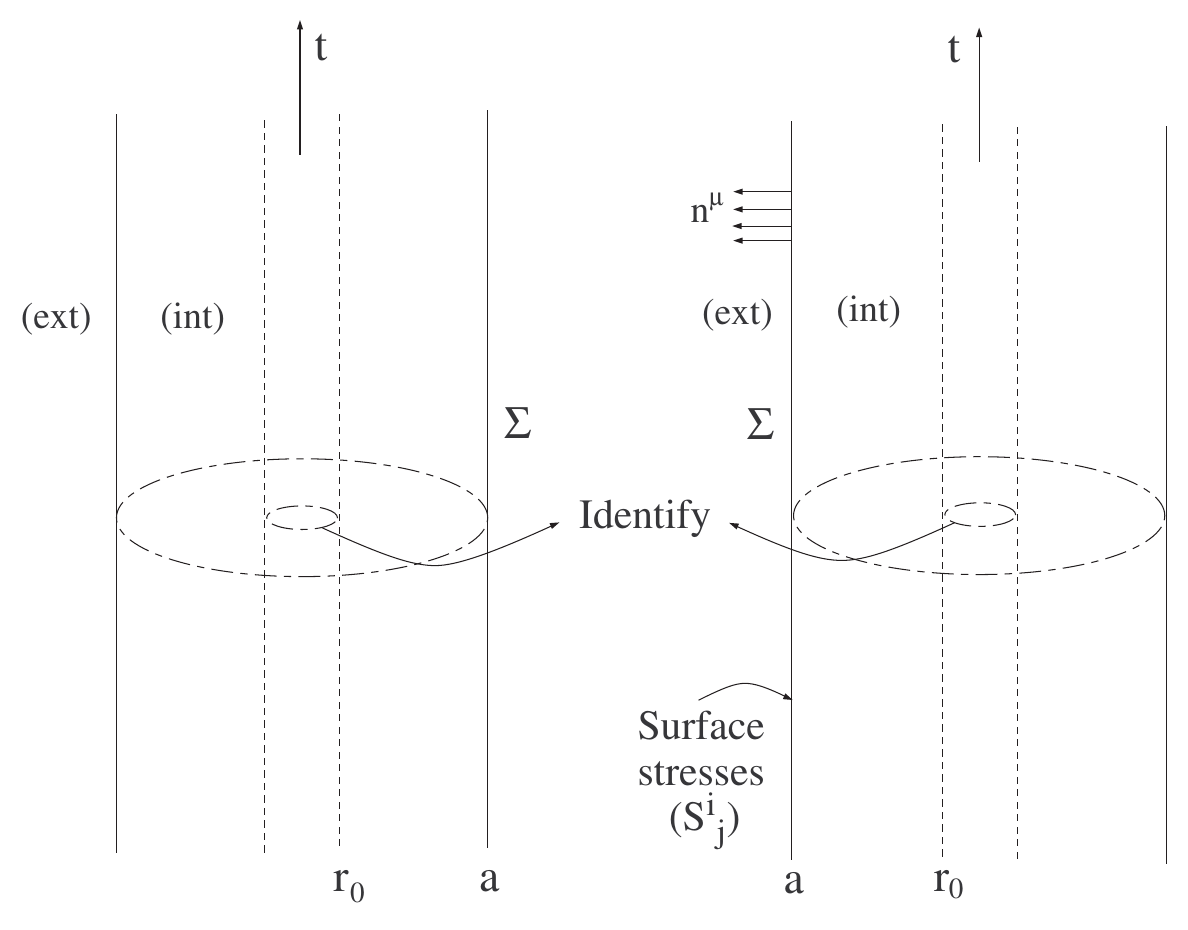}
	\caption{Two 
	copies of static timelike hypersurfaces, $\Sigma$, embedded in
		asymptotic regions, separating an interior wormhole solution from
		an exterior vacuum spacetime. Both copies are identified at the wormhole
		throat, $r_0$. The surface stresses reside on $\Sigma$, and members of the
		normal vector field, $n^{\mu}$, \mbox{are shown} (plot adapted from Figure 1 of Ref.~\cite{Lobo:2004id}).}
		\label{fig4}
\end{figure}

We shall match the interior wormhole solution~\cite{Morris:1988cz}
\begin{equation}
	ds^2=-e ^{2\Phi(r)}\,dt^2+\frac{dr^2}{1- b(r)/r}+r^2 \,(d\theta
	^2+\sin ^2{\theta} \, d\phi ^2)   ,
	\label{metricwormhole}
\end{equation}
to the exterior vacuum solution
\begin{eqnarray}
	ds^2=-\left(1-\frac{2M}{r}-\frac{\Lambda}{3}r^2
	\right)\,dt^2+\left(1-\frac{2M}{r}-\frac{\Lambda}{3}r^2
	\right)^{-1}\,dr^2 +r^2\,(d\theta ^2+\sin ^2{\theta}\, d\phi ^2)
	,
\end{eqnarray}
at a junction surface, $\Sigma$. Note that the interior solution, given by Equation (\ref{metricwormhole}), should also be matched, at the throat $r_0$, to two distinct ''upper'' and ``lower'' spacetimes, where the metric functions could assume different values, namely, $\Phi_{-}(r_-) \neq \Phi_{+}(r_+)$ and $b_{-}(r_-) \neq b_{+}(r_+)$. However, throughout this work we consider that the solutions are symmetric, so that $\Phi_{-}(r_-) = \Phi_{+}(r_+) = \Phi(r)$ and $b_{-}(r_-) = b_{+}(r_+)=b(r)$. See Figure \ref{fig4} for specific details, where the plot demonstrates the symmetric matching at the throat $r_0$, where both spacetimes are identified at the $r_0$ (ror asymmetric matchings, we refer the reader to Ref.~\cite{Garcia:2011aa}).
Thus, the intrinsic metric to
$\Sigma$ is given by
\begin{equation}
	ds^2_{\Sigma}=-d\tau^2 + a^2 \,(d\theta ^2+\sin
	^2{\theta}\,d\phi^2)  .
\end{equation}
Note that the junction surface, $r=a$, is situated outside the
event horizon, i.e., $a>r_b$, to avoid a black hole solution.

Thus, using Equation (\ref{extrinsiccurv}), the non-trivial
components of the extrinsic curvature are given by
\begin{eqnarray}
	K ^{\tau \;+}_{\;\;\tau}&=&\frac{\frac{M}{a^2}-
		\frac{\Lambda}{3}a}{\sqrt{1-\frac{2M}{a}-\frac{\Lambda}{3}a^2}}
	,  \label{Kplustautau2}\\
	K ^{\tau \;-}_{\;\;\tau}&=&\Phi'(a)\sqrt{1-\frac{b(a)}{a}}  ,
	\label{Kminustautau2}
\end{eqnarray}
and
\begin{eqnarray}
	K ^{\theta \;+}_{\;\;\theta}&=&\frac{1}{a}\sqrt{1-\frac{2M}{a}-
		\frac{\Lambda}{3}a^2},  \label{Kplustheta2}\\
	K ^{\theta \;-}_{\;\;\theta}&=&\frac{1}{a}\sqrt{1-\frac{b(a)}{a}}
	.  \label{Kminustheta2}
\end{eqnarray}

\textls[+25]{The Einstein equations, Equations
(\ref{sigma}) and (\ref{surfacepressure}), with the extrinsic
curvatures, Equations (\ref{Kplustautau2})--(\ref{Kminustheta2})},
then provide us with the following expressions for the surface
stresses:
\begin{eqnarray}
	\sigma&=&-\frac{1}{4\pi a} \left(\sqrt{1-\frac{2M}{a}-
		\frac{\Lambda}{3}a^2}- \sqrt{1-\frac{b(a)}{a}} \, \right)
	\label{surfenergy2}   ,\\
	p&=&\frac{1}{8\pi a} \left(\frac{1-\frac{M}{a}
		-\frac{2\Lambda}{3}a^2}{\sqrt{1-\frac{2M}{a}-\frac{\Lambda}{3}a^2}}-
	\zeta \, \sqrt{1-\frac{b(a)}{a}} \, \right)
	\label{surfpressure2}    ,
\end{eqnarray}
where, for notational simplicity, we have defined the parameter $\zeta$ by $\zeta=1+a\Phi'(a)$.

If the surface stress--energy terms are
null, the junction is denoted as a boundary surface. If surface
stress terms are present, the junction is called a thin shell,
which is represented in Figure \ref{fig4}.

The surface mass of the thin shell is given by
\begin{eqnarray}\label{shellmass}
	M_{\rm shell}=4\pi a^2 \sigma
	=a\left(\sqrt{1-\frac{b(a)}{a}}-\sqrt{1-\frac{2M}{a}-
		\frac{\Lambda}{3}a^2} \, \right)  .
\end{eqnarray}
One may interpret $M$ as the total mass of the system, in this
case, the total mass of the wormhole in one asymptotic
region. Thus, solving Equation (\ref{shellmass}) for $M$, we
finally have
\begin{equation}\label{totalmass}
	M=\frac{b(a)}{2}+M_{\rm
		shell}\left(\sqrt{1-\frac{b(a)}{a}}-\frac{M_{\rm
			shell}}{2a}\right)-\frac{\Lambda}{6}a^3   .
\end{equation}

In this section, we introduce several definitions to simplify the notation of the parameters used in the thin-shell formalism. For clarity, these dimensionless parameters are summarized in Table \ref{tab:parameters} to assist the reader during the analysis.

\begin{table}[H]
\caption{Definitions 
 of the dimensionless parameters used throughout this section for the solutions under analysis.}\label{tab:parameters}
 \begin{tabularx}{\textwidth}{CCc}
	\toprule
		\textbf{Schwarzschild} & \textbf{Schwarzschild--de Sitter} & \textbf{Schwarzschild--anti-de Sitter} \\
		\midrule
		$\zeta=1+a\Phi'(a)$ & $\zeta=1+a\Phi'(a)$ & $\zeta=1+a\Phi'(a)$ \\
		$\xi=2M/a$ & $\xi=2M/a$ & $\xi=2M/a$ \\
			 & $\beta=9 \Lambda M^2$ & $\gamma=9 |\Lambda| M^2$ \\
		$\mu=8\pi M\sigma$ & $\mu=8\pi M\sigma$ & $\mu=8\pi M\sigma$ \\
	\bottomrule
	\end{tabularx}
	
\end{table}

\subsection{Energy Conditions at the Junction Surface}

The junction surface may serve to confine the interior wormhole
exotic matter to a finite region, which in principle may be made
arbitrarily small, and one may impose that the surface
stress--energy tensor obeys the energy conditions at the junction,
$\Sigma$~\cite{Visser:1995cc,Hawking:1973uf}.

We shall only consider the weak energy condition (WEC) and the
null energy condition (NEC). The WEC implies $\sigma \geq 0$ and
$\sigma + p \geq 0$ and by continuity implies the null energy
condition (NEC), $\sigma + p\geq 0$.

From Equations (\ref{surfenergy2}) and (\ref{surfpressure2}), we deduce
\begin{equation} \label{sigma+P}
	\sigma +p=\frac{1}{8\pi a}
	\left[(2-\zeta)\,\sqrt{1-\frac{b(a)}{a}} -
	\frac{1-\frac{3M}{a}}{\sqrt{1-\frac{2M}{a}-\frac{\Lambda}{3}a^2}}
	\right]  .
\end{equation}
We shall next find domains in which the NEC is satisfied  by
imposing that the surface energy density is non-negative, $\sigma
\geq 0$, i.e., $\sqrt{1-b(a)/a} \geq \sqrt{1-2M/a-\Lambda a^2/3}$.

\subsubsection{Schwarzschild Solution}

Consider the Schwarzschild solution, $\Lambda=0$; we impose
that the surface energy density is non-negative, $\sigma \geq 0$.
For the particular case of $\zeta \leq 1$, from Equation
(\ref{sigma+P}), we verify that $\sigma+p\geq 0$ is readily
satisfied for $\forall \,a$.

For $1<\zeta <2$, the NEC is verified in the following region:
\begin{equation}\label{Schwregion}
	2M<a \leq 2M\,\left(\frac{\zeta-\frac{1}{2}}{\zeta-1}\right).
\end{equation}
For convenience, by defining a new parameter $\xi=2M/a$, Equation
(\ref{Schwregion}) takes the form
\begin{equation}\label{Schwregion2}
	\frac{\zeta-1}{\zeta-\frac{1}{2}} \leq \xi <1 .
\end{equation}

For $\zeta=2$, the NEC is satisfied for $\xi \geq 2/3$, i.e., $a
\leq 3M$. For $\zeta > 2$, we need to impose the NEC in the region
of Equation (\ref{Schwregion2}), with $\sigma+p<0$ for
$\xi<(\zeta-1)/(\zeta-1/2)$.

\subsubsection{Schwarzschild--de Sitter Solution}

For the Schwarzschild--de Sitter spacetime, $\Lambda > 0$, we shall
once again impose a non-negative surface energy density, $\sigma
\geq 0$. Consider the definitions $\beta=9 \Lambda M^2$ and
$\xi=2M/a$.

\begin{figure}[H]
	\includegraphics[width=1.9in]{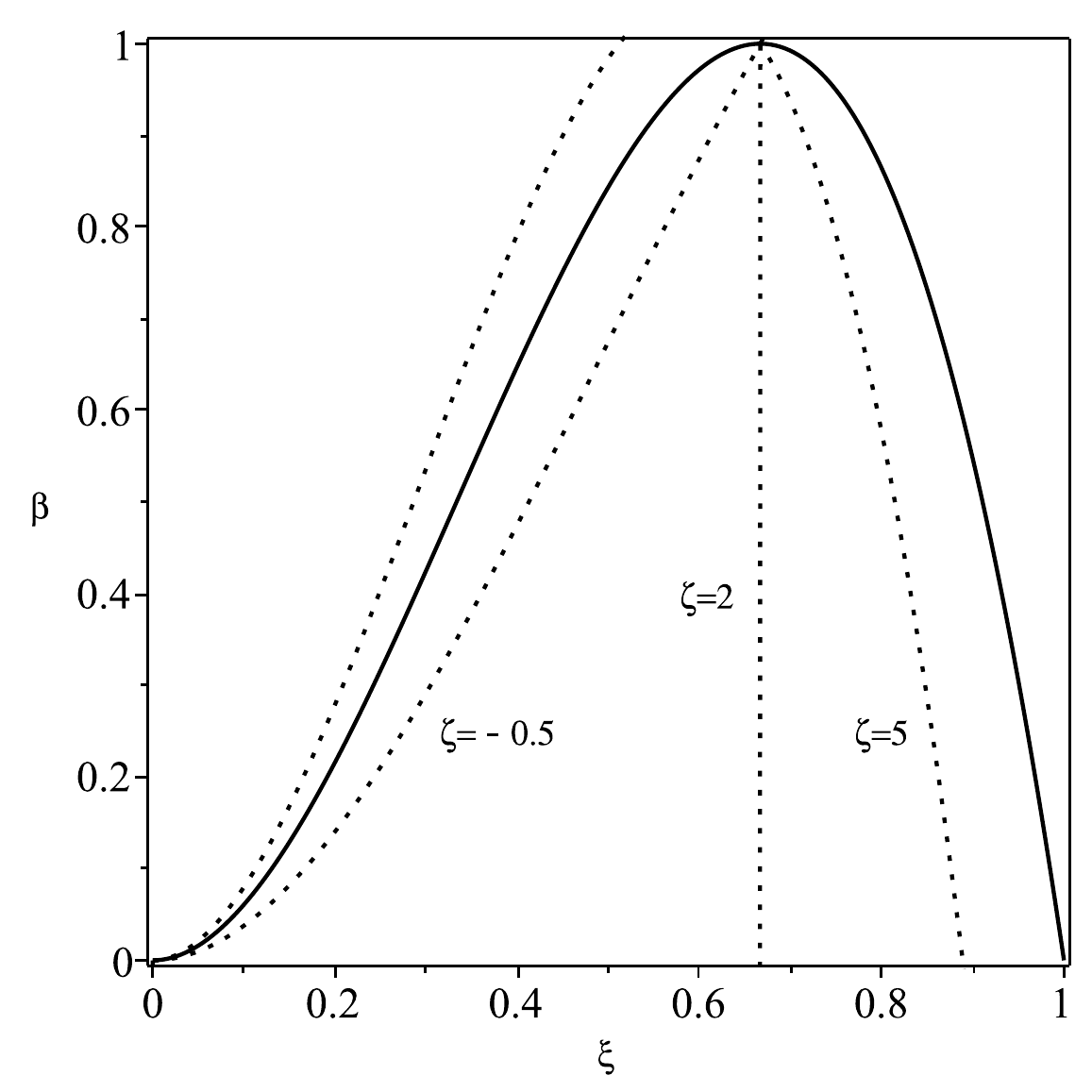}
	\caption{Analysis 
 of the null energy condition for the
		Schwarzschild--de Sitter spacetime. We have considered the
		definitions $\beta=9 \Lambda M^2$ and $\xi=2M/a$.
		Only the region below the solid line is of interest. We have
		considered specific examples, and the NEC is
		obeyed to the right of each respective dashed curve,
		$\zeta=-0.5$,  $\zeta=2$ and $\zeta=5$  (plot adapted from Figure 4 of Ref.~\cite{Lobo:2004id}).
		See text for details.}
		\label{fig5}
\end{figure}

For $\zeta < 2$, the condition $\sigma+p\geq 0$ is readily met for
$\beta \leq \beta_0$, with $\beta_0$ given by
\begin{equation}\label{SdSbeta0}
	\beta_0=\frac{27}{4}\,\frac{\xi^2}{(2-\zeta)}\,\left[(1-\zeta)
	+\left(\zeta-\frac{1}{2}\right)\xi \right].
\end{equation}
Choosing a particular example, for instance, $\zeta=-0.5$, consider
Figure \ref{fig5}. The region of interest is shown below the solid line,
which is given by $\beta_r=27\xi^2(1-\xi)/4$. The case of
$\zeta=-0.5$ is depicted as a dashed curve, and the NEC is obeyed
to the right of the latter.

For $\zeta=2$, then, the NEC is verified for $\forall \,\beta$ and
$\xi \geq 2/3$, i.e., $r_b<a \leq 3M$, with $r_b$ given by
Equation (\ref{root1}). This analysis is depicted in Figure \ref{fig5}, to
the right of the dashed curve, represented by $\xi=2$.

For the case of $\zeta>2$, the condition $\sigma+p\geq 0$ needs to
be imposed in the region $\beta_0 \leq \beta \leq \beta_r$; and
$\sigma+p< 0$ for $\beta < \beta_0$. The specific case of
$\zeta=5$ is depicted as a dashed curve in Figure \ref{fig5}. The NEC needs
to be imposed to the right of the respective curve.

\subsubsection{Schwarzschild--anti-de Sitter Solution}

Considering the Schwarzschild--anti-de Sitter spacetime, $\Lambda <
0$, once again, a non-negative surface energy density, $\sigma \geq
0$, is imposed. Consider the definitions $\gamma=9 |\Lambda| M^2$
and $\xi=2M/a$.

\begin{figure}[H]
	\includegraphics[width=1.9in]{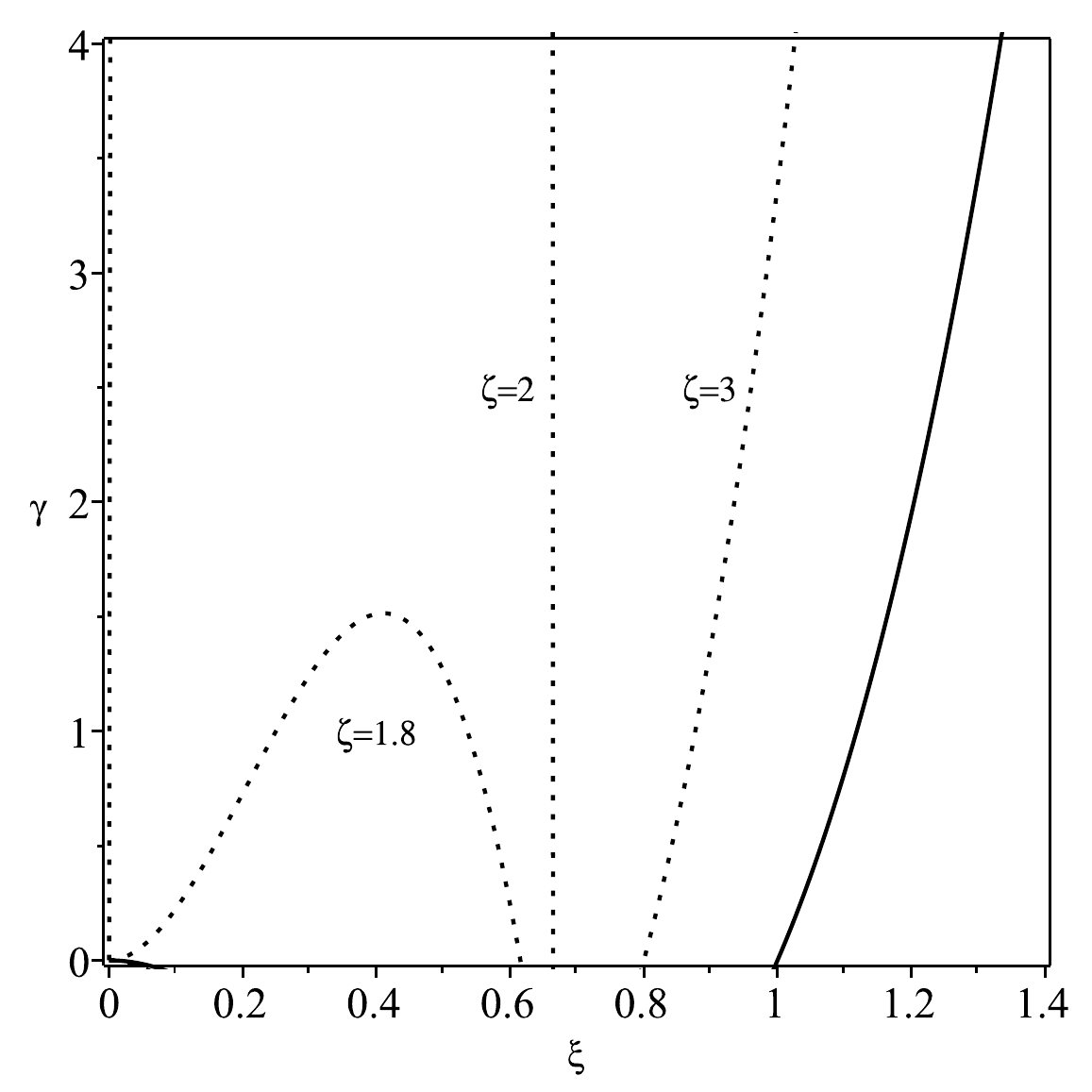}
	\caption{Analysis of the null energy condition for the
		Schwarzschild--anti-de Sitter spacetime. We have considered the
		definitions $\gamma=9 |\Lambda| M^2$ and $\xi=2M/a$.
		The only area of interest is depicted to the left of the solid curve,
		given by $\gamma_r=27\xi^2(\xi-1)/4$. For the specific case of
		$\zeta=1.8$, the NEC is obeyed above the respective curve. For
		the cases of $\zeta=2$ and $\zeta=3$, the NEC is verified to the
		right of the respective dashed curves, and to the left of the
		solid line  (plot adapted from Figure 5 of Ref.~\cite{Lobo:2004id}). See text for details.}
		\label{fig6}
\end{figure}

For $\zeta \leq 1$, the condition $\sigma+p\geq 0$ is readily met
for $\forall \,\gamma$ and $\forall \,\xi$. For $1<\zeta<2$, the
NEC is satisfied in the region $\gamma \geq \gamma_0$, with
$\gamma_0$ given by
\begin{equation}\label{SadSbeta0}
	\gamma_0=\frac{27}{4}\,\frac{\xi^2}{(2-\zeta)}\,\left[(1-\zeta)
	+\left(\zeta-\frac{1}{2} \right)\xi  \right].
\end{equation}
The particular case of $\zeta=1.8$ is depicted in Figure \ref{fig6}. The
region of interest is delimited  by the $\xi$-axis and the area to
the left of the solid curve, which is given by
$\gamma_r=27\xi^2(\xi-1)/4$. Thus, the NEC is obeyed above the
dashed curve represented by the value $\zeta=1.8$.

For $\zeta=2$, then, $\sigma+p\geq 0$ is verified for $\forall
\gamma$ and $\xi \geq 3/2$, i.e., $r_b<a \leq 3M$, with $r_b$
given by Equation (\ref{adsbhole}). Thus, the NEC is obeyed to the
right of the dashed curve represented by $\zeta=2$, and to the
left of the solid line, $\gamma_r$.

For the case of $\zeta>2$, the condition $\sigma+p\geq 0$ needs to
be imposed in the region $\gamma_r \leq \gamma \leq \gamma_0$. The
specific case of $\zeta=3$ is depicted in Figure \ref{fig6} as a dashed
curve. Thus, the NEC needs to be imposed in the region to the
right of the respective dashed curve and to the left of the solid
line, $\gamma_r$.

\subsection{Specific Cases}

Taking into account Equations
(\ref{surfenergy2}) and (\ref{surfpressure2}), one may express $p$ as
a function of $\sigma$ by the following relationship
\begin{equation}
	p=\frac{1}{8\pi a}
	\,\left[\frac{(1-\zeta)+(\zeta-\frac{1}{2})\,\frac{2M}{a}
		-(2-\zeta)\frac{\Lambda}{3}a^2}{\sqrt{1-\frac{2M}{a}-
			\frac{\Lambda}{3}a^2}} -4\pi a \zeta \sigma \right] .
	\label{Pfunctionsigma}
\end{equation}
We shall analyze Equation (\ref{Pfunctionsigma}), that is, find
domains in which $p$ assumes the nature of a tangential surface
pressure, $p>0$, or a tangential surface tension, $p<0$, for the
Schwarzschild case, $\Lambda=0$, the Schwarzschild--de Sitter
spacetime, $\Lambda>0$, and for the Schwarzschild--anti-de Sitter
solution, $\Lambda<0$. In the analysis that follows, we shall
consider that $M$ is positive, $M>0$.

\subsubsection{Schwarzschild Spacetime}

For the Schwarzschild spacetime, $\Lambda=0$, Equation
(\ref{Pfunctionsigma}) reduces to
\begin{equation}
	p=\frac{1}{8\pi a}
	\,\left[\frac{(1-\zeta)+(\zeta-\frac{1}{2})\,\frac{2M}{a}}{\sqrt{1-\frac{2M}{a}}}
	-4\pi a \zeta \sigma \right].
	\label{SchwarzPfunctionsigma}
\end{equation}
To find domains in which $p$ is a tangential surface pressure,
$p>0$, or a tangential surface tension, $p<0$, it is convenient to
express Equation (\ref{SchwarzPfunctionsigma}) in the  
compact form
\begin{equation}
	p=\frac{1}{16\pi M} \,
	\frac{\Gamma(\xi,\zeta,\mu)}{\sqrt{1-\xi}},
	\label{SchwarzcompactP}
\end{equation}
with $\xi=2M/a$ and $\mu=8\pi M\sigma$. $\Gamma(\xi,\zeta,\mu)$ is
defined as
\begin{eqnarray}\label{SchwarzGamma}
	\Gamma(\xi,\zeta,\mu)=(1-\zeta)\,\xi+\left(\zeta-\frac{1}{2}\right)\xi^2-\mu
	\zeta \sqrt{1-\xi}    .
\end{eqnarray}
One may now fix one or several of the parameters and analyze the
sign of $\Gamma(\xi,\zeta,\mu)$, and consequently the sign of $p$.

\bigskip

{\it Fixed 
 $\zeta$, varying $\xi$ and $\mu$.} For instance,
consider a fixed value of $\zeta$, varying the parameters
$(\xi,\mu)$; i.e., consider a fixed value of $a\Phi'(a)$ and vary
the values of the junction radius, $a$, and of the surface energy
density, $\sigma$. It is necessary to separate the cases of
$\zeta=0$, $\zeta>0$ and $\zeta<0$, respectively.

Firstly, for the case of $\zeta=0$, Equation (\ref{SchwarzGamma})
reduces to $\Gamma(\xi,\zeta=0,\mu)=\xi-\xi^2/2$, which is always
positive, as $0<\xi<1$, implying a tangential surface pressure,
$p>0$.

\begin{figure}[H]
	\includegraphics[width=2.6in]{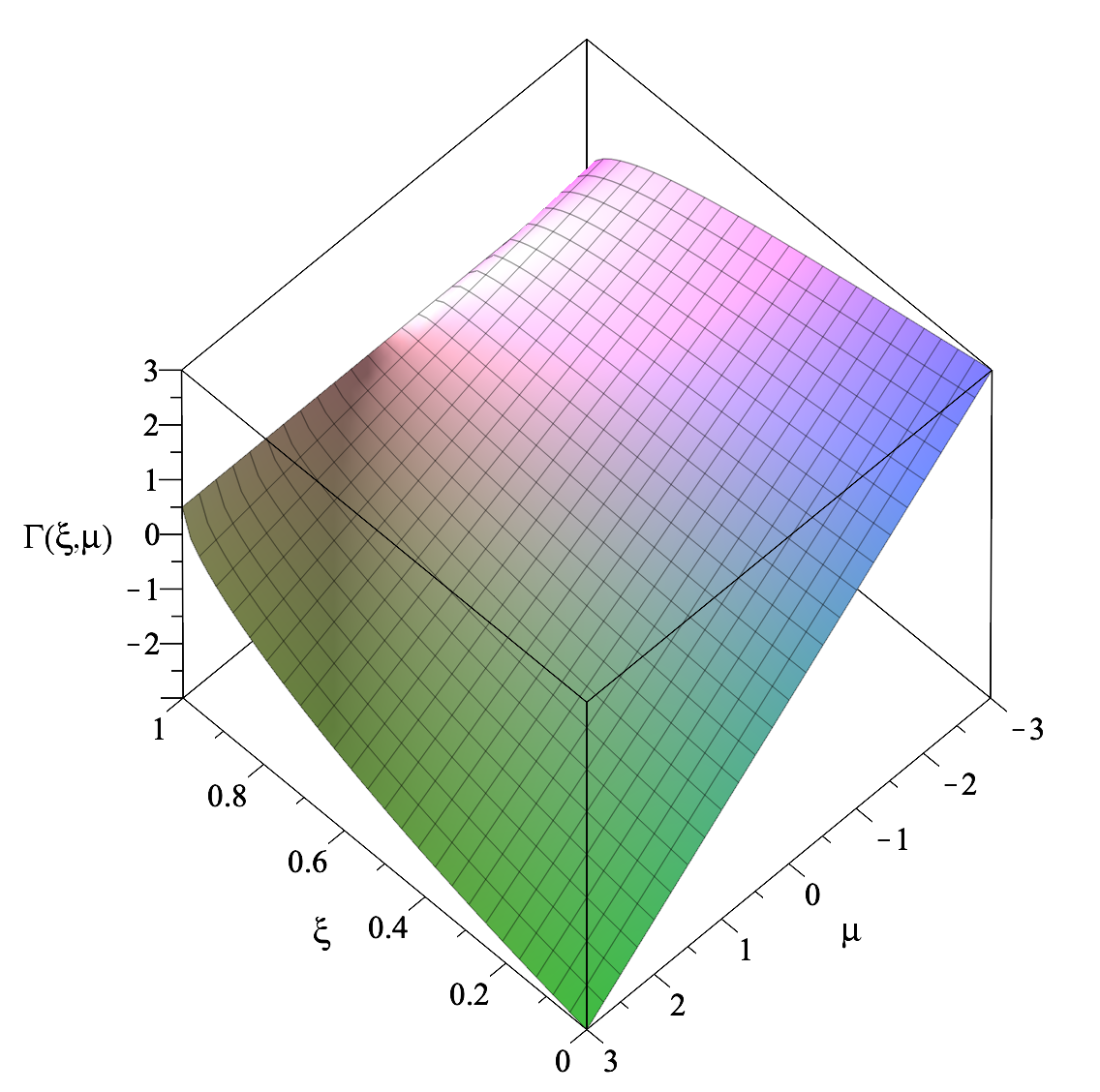}
	\caption{\textls[-15]{The surface represents the sign of $p$ for the
		Schwarzschild spacetime, $\Lambda=0$, with a
		constant redshift function, $\Phi'(r)=0$, i.e., $\zeta=1$.
		For non-positive values of $\mu$ and $\forall \,\xi$,
		we have a surface tangential}}
		\label{fig7}
\end{figure}

Secondly, for $\zeta>0$, the qualitative behavior can be
represented by the specific case of $\zeta=1$, corresponding to a
constant redshift function and depicted in Figure \ref{fig7}. For
non-positive values of $\mu$ and $\forall \,\xi$, a surface
pressure, $p>0$, is required to hold the thin shell structure
against collapse. Close to the black hole event horizon, $a
\rightarrow 2M$, i.e., $\xi \rightarrow 1$, a surface pressure is
also needed to hold the structure against collapse. For high
values of $\mu$ and low values of $\xi$, a surface tangential
tension, $p<0$, is needed to hold the structure against expansion.
In particular, for the constant redshift function, $\Phi'(r)=0$,
and a null surface energy density, $\sigma=0$, i.e., $\zeta=1$ and
$\mu=0$, respectively, Equation (\ref{SchwarzGamma}) reduces to
$\Gamma(\xi)=\xi^2/2$, from which we readily conclude that $p$ is
non-negative everywhere, tending to zero at infinity, i.e., $\xi
\rightarrow 0$. This is a particular case analyzed in~\cite{Lemos:2003jb}.
Note that a surface boundary, with $p=0$ and $\sigma=0$, is given
by $\xi=(\zeta-1)/(\zeta-1/2)$, for $\zeta>1$.

\begin{figure}[H]
	\includegraphics[width=2.7in]{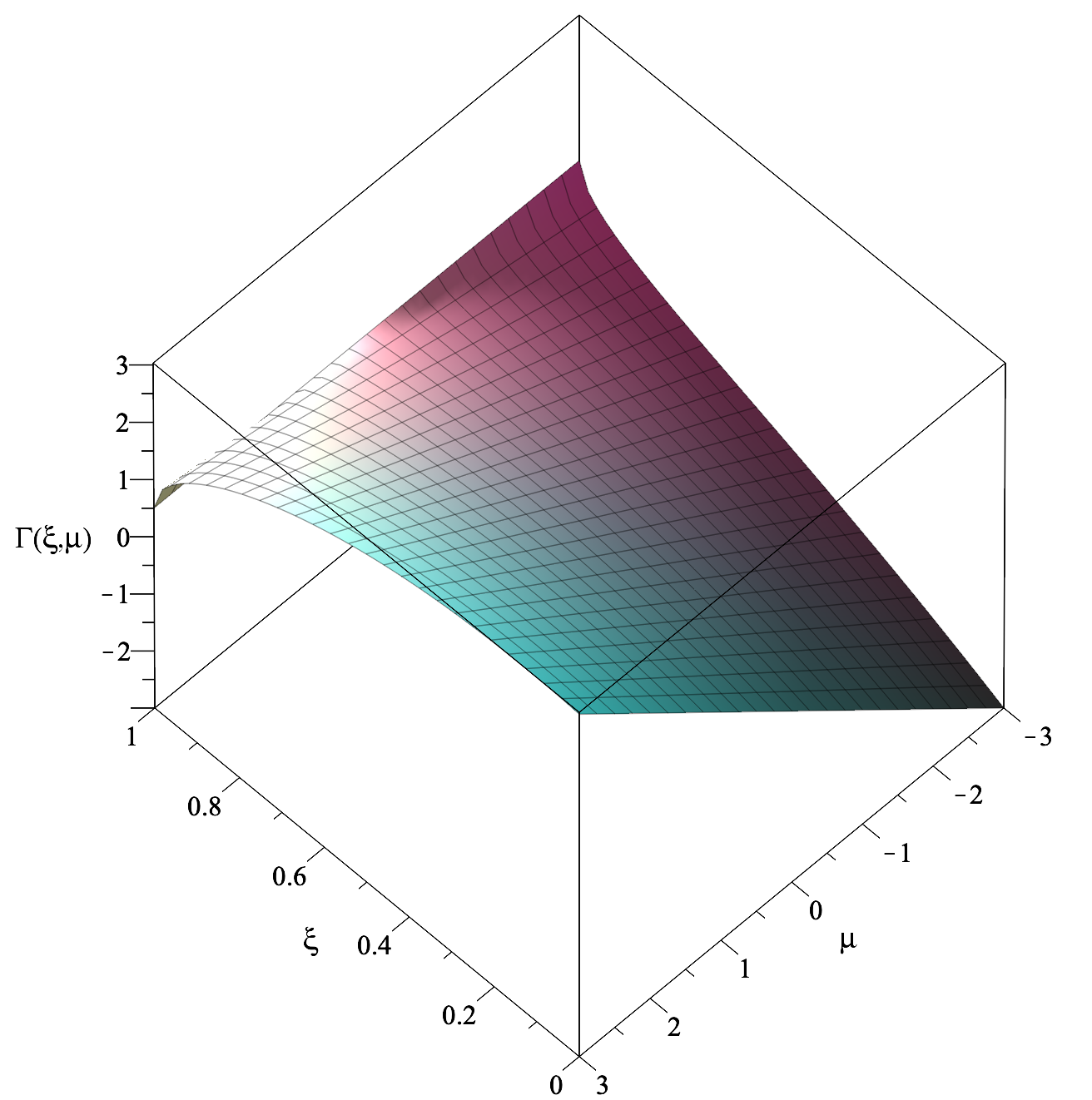}
	\caption{The surface is given by Equation (\ref{SchwarzGamma}) for the
		Schwarzschild spacetime, $\Lambda=0$, with $\zeta=-1$.
		For non-negative values of $\mu$ and $\forall \,\xi$,
		we have a surface tangential pressure, $p>0$.
		For negative values of $\mu$ and low $\xi$, a
		tangential surface tension, $p<0$, is required to hold
		the structure against expansion. See text for details.}
		\label{fig8}
\end{figure}

\textls[-15]{Finally, for the $\zeta<0$ case, the qualitative behavior can be
represented by the specific case of $\zeta=-1$, depicted in Figure
\ref{fig8}. For non-negative values of $\mu$ and for $\forall \,\xi$, a
surface pressure, $p>0$, is required. For low negative values of
$\mu$ and for low values of $\xi$, a surface tension is needed,
which is somewhat intuitive as a negative surface energy density
is gravitationally repulsive, requiring a surface tension to hold
the structure \mbox{against expansion}.}

\vspace{-3pt}
\vspace{-12pt}
{\captionof*{figure}{
		pressure, $p>0$. For extremely high values of $\xi$
		(close to the black hole event horizon) and $\forall \,\mu$,
		a surface pressure is also required to hold the structure
		against collapse. For high $\mu$ and low $\xi$,
		we have a
		tangential surface tension, $p<0$. See text for details.}}
\vspace{12pt}

\vspace{-6pt}

\subsubsection{Schwarzschild--de Sitter Spacetime}

For the Schwarzschild--de Sitter spacetime with $\Lambda>0$, to
analyze the sign of $p$, it is convenient to express
Equation (\ref{Pfunctionsigma}) in the following compact form:
\begin{equation}
	p=\frac{1}{16\pi M} \,
	\frac{\Gamma(\xi,\zeta,\mu,\beta)}{\sqrt{1-\xi-\frac{4\beta}{27\xi^2}}},
	\label{compactP}
\end{equation}
with $\xi=2M/a$, $\mu=8\pi M\sigma$ and $\beta=9\Lambda M^2$.
$\Gamma(\xi,\zeta,\mu,\beta)$ is defined as
\begin{eqnarray}\label{GammadS}
	\Gamma(\xi,\zeta,\mu,\beta)=(1-\zeta)\,\xi+\left(\zeta-\frac{1}{2}\right)\xi^2
	-\frac{4\beta}{27\xi}\,(2-\zeta) -\mu \zeta
	\sqrt{1-\xi-\frac{4\beta}{27\xi^2}}    .
\end{eqnarray}
One may now fix several of the parameters and analyze the sign of
$\Gamma(\xi,\zeta,\mu,\beta)$, and consequently the sign of $p$.

\bigskip

{\it Null surface energy density.} For instance, consider a null
surface energy density, $\sigma=0$, i.e., $\mu=0$. Thus, Equation
(\ref{GammadS}) reduces to
\begin{equation}\label{GammasigmadS}
	\Gamma(\xi,\zeta,\beta)=(1-\zeta)\,\xi+\left(\zeta-\frac{1}{2}\right)\,\xi^2-
	(2-\zeta)\,\frac{4\beta}{27\xi}.
\end{equation}
To analyze the sign of $p$, we shall consider a null tangential
surface pressure, i.e., \mbox{$\Gamma(\xi,\zeta,\beta)=0$}, so that from
Equation (\ref{GammasigmadS}), we have the   relationship
\begin{equation}\label{beta0}
	\beta_0=\frac{27}{4}\frac{\xi^2}{(2-\zeta)}\left[(1-\zeta)
	+\left(\zeta-\frac{1}{2}\right)\,\xi \right]  ,
\end{equation}
with $\zeta \neq 2$, which is identical to Equation
(\ref{SdSbeta0}).

For the particular case of $\zeta=2$, from Equation
(\ref{GammasigmadS}), we have
$\Gamma(\xi,\zeta=2,\beta)=\xi(3\xi/2-1)$. A surface boundary,
$\Gamma(\xi,\zeta=2,\beta)=0$, is presented at $\xi=2/3$; i.e.,
$a=3M$. A surface pressure, $\Gamma(\xi,\zeta=2,\beta)>0$, is
given for $\xi>2/3$, i.e., $r_b<a<3M$, and a surface tension,
$\Gamma(\xi,\zeta=2,\beta)<0$, for $\xi<2/3$, i.e., $3M<a<r_c$.

For $\zeta<2$, from Equation (\ref{GammasigmadS}), a surface
pressure, $\Gamma(\xi,\zeta,\beta)>0$, is met for $\beta<\beta_0$,
and a surface tension, $\Gamma(\xi,\zeta,\beta)<0$, for
$\beta>\beta_0$. The specific case of a constant redshift
function, i.e., $\zeta=1$, is analyzed in~\cite{Lemos:2003jb} (for this
case, Equation (\ref{beta0}) is reduced to $\beta_0=27\xi^3/8$, or
$M=\Lambda a^3/3$). For the qualitative behavior, the reader is
referred to the particular case of $\zeta=-0.5$ provided in Figure
\ref{fig5}. To the right of the curve a surface pressure, $p>0$, is given,
and to the left of the respective curve a surface tension, $p<0$.

For $\zeta>2$, from Equation (\ref{GammasigmadS}), a surface
pressure, $\Gamma(\xi,\zeta,\beta)>0$, is given for
\mbox{$\beta_0<\beta<\beta_r$}, and a surface tension,
$\Gamma(\xi,\zeta,\beta)<0$, for $\beta<\beta_0$. Once again, the
reader is referred to Figure \ref{fig5} for a qualitative analysis of the
behavior for the particular case of $\zeta=5$. A surface pressure
is given to the right of the curve and a surface tension to the
left.

Note that for the analysis considered in this section, namely, for
a null surface energy density, the WEC  and consequently the NEC 
are satisfied only if $p \geq 0$. The results obtained are
consistent with those of the section regarding the energy
conditions at the junction surface, for the Schwarzschild-de
Sitter spacetime, considered above.

\subsubsection{Schwarzschild--anti-de Sitter Spacetime}

For the Schwarzschild--anti-de Sitter spacetime, $\Lambda<0$, and to
analyze the sign of $p$, Equation (\ref{Pfunctionsigma}) is
expressed as
\begin{equation}
	p=\frac{1}{16\pi M} \,
	\frac{\Gamma(\xi,\zeta,\mu,\gamma)}{\sqrt{1-\xi+\frac{4\gamma}{27\xi^2}}},
	\label{SadScompactP}
\end{equation}
with the parameters given $\xi=2M/a$, $\mu=8\pi M\sigma$ and
$\gamma=9|\Lambda| M^2$, respectively.
$\Gamma(\xi,\zeta,\mu,\gamma)$ is defined as
\begin{eqnarray}\label{GammaSadS}
	\Gamma(\xi,\zeta,\mu,\gamma)=(1-\zeta)\,\xi+\left(\zeta-\frac{1}{2}\right)\xi^2
	+\frac{4\gamma}{27\xi}\,(2-\zeta) -\mu \zeta
	\sqrt{1-\xi+\frac{4\gamma}{27\xi^2}}    .
\end{eqnarray}
As in the Schwarzschild--de Sitter solution, we shall analyze the
case of a null surface energy density, $\sigma=0$, i.e., $\mu=0$.

{\it Null surface energy density.} \textls[+15]{For a null surface energy
density, $\sigma=0$, i.e., $\mu=0$, \mbox{Equation (\ref{GammaSadS})}
reduces to}
\begin{equation}\label{GammasigmaSadS}
	\Gamma(\xi,\zeta,\gamma)=(1-\zeta)\,\xi+\left(\zeta-\frac{1}{2}\right)\,\xi^2+
	(2-\zeta)\,\frac{4\gamma}{27\xi},
\end{equation}
To analyze the sign of $p$, once again, we shall consider a null
tangential surface pressure, i.e., $\Gamma(\xi,\zeta,\beta)=0$, so
that from Equation (\ref{GammasigmaSadS}), we have
\begin{equation}\label{SadSbeta}
	\gamma_0=\frac{27}{4}\frac{\xi^2}{(2-\zeta)}\left[(\zeta-1)
	-\left(\zeta-\frac{1}{2}\right)\,\xi \right]  ,
\end{equation}
with $\zeta \neq 2$, which is identical to Equation
(\ref{SadSbeta0}).

\textls[-15]{For the particular case of $\zeta=2$, from Equation
(\ref{GammasigmaSadS}), we have
\mbox{$\Gamma(\xi,\zeta=2,\beta)=\xi(3\xi/2-1)$,}} which is null at
$\xi=2/3$, i.e., $a=3M$. A surface pressure,
$\Gamma(\xi,\zeta=2,\gamma)>0$, is given for $\xi>2/3$, i.e.,
$r_b<a<3M$, and a surface tension, $\Gamma(\xi,\zeta=2,\beta)<0$,
for $\xi<2/3$, i.e., $a>3M$. The reader is referred to the
particular case of $\zeta=2$, depicted in Figure \ref{fig6}. A surface
pressure is given to the right of the respective dashed curve, and
a surface tension to the left.

For $\zeta \leq 1$, a surface pressure,
$\Gamma(\xi,\zeta,\gamma)>0$, is given for $\forall \; \gamma$ and
$\forall \;\xi$. For \mbox{$1<\zeta<2$}, a surface pressure,
$\Gamma(\xi,\zeta,\gamma)>0$, is given for $\gamma>\gamma_0>0$;
and a surface tension, \mbox{$\Gamma(\xi,\zeta,\gamma)<0$}, is provided
for $0<\gamma<\gamma_0$. The particular case of $\zeta=1.8$ is
depicted in Figure \ref{fig6}, in which a surface pressure is presented
above the respective dashed curve, and a surface tension is
presented in the region delimited by the curve and the $\xi$-axis.

For $\zeta>2$, a surface pressure, $\Gamma(\xi,\zeta,\gamma)>0$,
is met for $\gamma_r<\gamma<\gamma_0$, and a surface tension,
$\Gamma(\xi,\zeta,\gamma)<0$, for $\gamma>\gamma_0$. The specific
case for $\zeta=3$ is depicted in Figure \ref{fig6}. The surface pressure is
presented to the right of the respective curve  and the surface
tension to the left.

Once again, the analysis considered in this section is consistent
with the results obtained in the section regarding the energy
conditions at the junction surface, for the Schwarzschild-anti de
Sitter spacetime, considered above. This is due to the fact that
for the specific case of a null surface energy density, the
regions in which the WEC and NEC are satisfied coincide with the
range of $p \geq 0$.

\subsection{Pressure Balance Equation}

One may obtain an equation governing the behavior of the radial
pressure in terms of the surface stresses at the junction boundary
from the following identity~\cite{Visser:1995cc,Musgrave:1995ka}:
$\left[\,T^{\rm
	total}_{\hat{\mu}\hat{\nu}}\,n^{\hat{\mu}}n^{\hat{\nu}}
\right]=\frac{1}{2}(K^{i\;+}_{\;\,j} +
K^{i\;-}_{\;\,j})\,S^{j}_{\;\,i}$, where $T^{\rm
	total}_{\hat{\mu}\hat{\nu}}=T_{\hat{\mu}\hat{\nu}}-g_{\hat{\mu}\hat{\nu}}\,\Lambda/8\pi$
is the total stress--energy tensor, and the square brackets denotes
the discontinuity across the thin shell, i.e.,
$[X]=X^{+}|_{\Sigma}-X^{-}|_{\Sigma}$. Taking into account the
values of the extrinsic curvatures, \mbox{Equations
(\ref{Kplustautau2})--(\ref{Kminustheta2})}, and noting that the
tension acting on the shell is by definition the normal component
of the stress--energy tensor,
$-\tau=T_{\hat{\mu}\hat{\nu}}\,n^{\hat{\mu}}n^{\hat{\nu}}$, we
finally have the following pressure balance equation 

\begin{eqnarray}\label{pressurebalance}
	\left(-\tau^+(a)-\frac{\Lambda ^+}{8\pi} \right) - \left(-\tau
	^-(a)-\frac{\Lambda ^-}{8\pi} \right)=
	\frac{1}{a}\,\left(\sqrt{1-\frac{2M}{a}-\frac{\Lambda}{3}a^2}
	+\sqrt{1-\frac{b(a)}{a}}\;\right)\,p
	\nonumber       \\
	-\left(\frac{\frac{M}{a^2}- \frac{\Lambda}{3}a}
	{\sqrt{1-\frac{2M}{a}-\frac{\Lambda}{3}a^2}}+\Phi'(a)\,\sqrt{1-\frac{b(a)}{a}}
	\right) \frac{\sigma}{2} ,
\end{eqnarray}
where the $\pm$ superscripts correspond to the exterior and
interior spacetimes, respectively. Equation
(\ref{pressurebalance}) relates the difference in the radial
tension across the shell in terms of a combination of the surface
stresses, $\sigma$ and $p$, given by Equations
(\ref{surfenergy2}) and (\ref{surfpressure2}), respectively, and the
geometrical quantities.

Note that for the exterior vacuum solution, we have $\tau^+=0$. For
the particular case of a null surface energy density, $\sigma=0$,
and considering that the interior and exterior cosmological
constants are equal, $\Lambda^-=\Lambda^+$, Equation
(\ref{pressurebalance}) reduces to
\begin{equation}
	\tau
	^-(a)=\frac{2}{a}\,\sqrt{1-\frac{2M}{a}-\frac{\Lambda}{3}a^2}\;\,p
	\,.
\end{equation}
For a radial tension, $\tau^-(a)>0$, acting on the shell from the
interior, a tangential surface pressure, $p>0$, is needed to hold
the collapsing thin shell form
. For a radial interior pressure,
$\tau^-(a)<0$, then, a tangential surface tension, $p<0$, is needed
to hold the structure form expansion.

\subsection{Traversability Conditions}

In this section, we shall consider the traversability conditions
required for the traversal of a human being through the wormhole
and consequently determine specific dimensions for the wormhole.
Specific cases for the traversal time and velocity will also be
estimated.

Consider the redshift function given by $\Phi(r)=kr^{\alpha}$,
with $\alpha, k\in \mathbb{R}$. Thus, from the definition of
$\zeta=1+a\Phi'(a)$, the redshift function, in terms of $\zeta$,
takes the   form
\begin{equation}\label{redshift}
	\Phi(r)=\frac{\zeta-1}{\alpha}\,\left(\frac{r}{a}\right)^{\alpha}
	,
\end{equation}
with $\alpha \neq 0$. With this choice of $\Phi(r)$, $\zeta$ may
also be defined as $\zeta=1+\alpha \Phi(a)$. The case of
$\alpha=0$ corresponds to the constant redshift function so that
$\zeta=1$. If $\alpha<0$, then $\Phi(r)$ is finite throughout
spacetime and in the absence of an exterior solution, we have
$\lim_{r\rightarrow \infty} \Phi(r)\rightarrow 0$. As we are
considering a matching of an interior solution with an exterior
solution at $a$, then it is also possible to consider the
$\alpha>0$ case, imposing that $\Phi(r)$ is finite in the interval
$r_0 \leq r \leq a$.

In the following analysis, we consider a thought experiment in which a
traveler journeys radially through a wormhole. Thus, the analysis outlined below is valid for a radially moving observer.
One of the traversability conditions was that the acceleration
felt by the traveler should not exceed Earth's gravity
\cite{Morris:1988cz}. Consider an orthonormal basis of the traveler's
proper reference frame, $({\bf e}_{\hat{0}'},{\bf
	e}_{\hat{1}'},{\bf e}_{\hat{2}'},{\bf e}_{\hat{3}'})$ 
, given in
terms of the orthonormal basis of the static observers, by a
Lorentz transformation. The traveler's four-acceleration
expressed in his proper reference frame,
$a^{\hat{\mu}'}=U^{\hat{\nu}'} U^{\hat{\mu}'}_{\;\;\;;
	\hat{\nu}'}$, yields the following restriction
\begin{equation}\label{travellergravity}
	\Bigg| \left(1-\frac{b}{r} \right)^{1/2} \;e^{-\Phi}\,(\gamma
	e^{\Phi})'\,c^2 \Bigg| \leq g_{\oplus}  ,
\end{equation}
where $\gamma=(1-v^2)^{-1/2}$, and $v$ is the velocity of the
traveler~\cite{Morris:1988cz}. The condition is immediately satisfied at
the throat, $r_0$. From Equation (\ref{travellergravity}), one may
also find an estimate for the junction surface, $a$. Considering
that $(1-b(a)/a)^{1/2}\approx 1$ and $\gamma \approx 1$, i.e., for
low traversal velocities, and taking into account Equation
(\ref{redshift}), from Equation (\ref{travellergravity}), one
deduces $a \geq |\zeta-1|c^2/g_{\oplus}$. Considering the equality
case, one has
\begin{equation}\label{equalitycase2}
	a = \frac{|\zeta-1|c^2}{g_{\oplus}}  .
\end{equation}
Providing a value for $|\zeta-1|$, one may find an estimate for
$a$. For instance, considering that $|\zeta-1|\simeq 10^{-10}$,
one finds that $a \approx 10^6 \,{\rm m}$.

Another of the traversability conditions that was required  was
that the tidal accelerations felt by the traveler should not
exceed the Earth's gravitational acceleration. The tidal
acceleration felt by the traveler is given by $\Delta
a^{\hat{\mu}'}=-R^{\hat{\mu}'}_{\;\;\hat{\nu}'\hat{\alpha}'\hat{\beta}'}
\,U^{\hat{\nu}'}\eta^{\hat{\alpha}'}U^{\hat{\beta}'}$, where
$U^{\hat{\mu}'}$ is the traveler's four-velocity and
$\eta^{\hat{\alpha}'}$ is the separation between two arbitrary
parts of his body. For simplicity, we shall assume that
$|\eta^{\hat{\alpha}'}|\approx 2\,{\rm m}$ along any spatial
direction in the traveler's reference frame, and that $|\Delta
a^{\hat{\mu}'}|\leq g_{\oplus}$. Consider the radial tidal
constraint,
$|R_{\hat{1}'\hat{0}'\hat{1}'\hat{0}'}|=|R_{\hat{r}\hat{t}\hat{r}\hat{t}}|
\leq g_{\oplus}/2c^2$, which can be regarded as constraining the
metric field \mbox{$\Phi(r)$, i.e.,}
\begin{equation}\label{radialtidalconstraint}
	\Bigg|\left(1-\frac{b}{r}\right)
	\left(-\Phi''+\frac{b'r-b}{2r(r-b)}\,\Phi'-(\Phi')^2 \right)\Bigg|
	\leq \frac{g_{\oplus}}{2c^2}  .
\end{equation}
At the throat, $r=r_0$, and taking into account Equation
(\ref{redshift}), then Equation (\ref{radialtidalconstraint})
reduces to $|(b'-1)\Phi'(r_0)/2r_0| \leq g_{\oplus}/2c^2$ or
\begin{equation}\label{equalitycase}
	a =
	\left(\frac{|b'-1|\,|\zeta-1|c^2}{g_{\oplus}r_0^2}\right)^{1/\alpha}\;r_0
	.
\end{equation}
considering the equality case.

Using Equations (\ref{equalitycase2}) and (\ref{equalitycase}), one may
find an estimate for the throat radius by providing a specific
value for $\alpha$. Considering $\alpha=-1$ and equating Equations
(\ref{equalitycase2}) and (\ref{equalitycase}), one finds
\begin{equation}\label{r0}
	r_0=\left(\frac{|b'-1|\,|\zeta-1|^2\,c^4}{g_{\oplus}^2}\right)^{1/3}
	.
\end{equation}
Kuhfittig~\cite{Kuhfittig:1999nd,Kuhfittig:2002ur,Kuhfittig:2003pu} proposed models
restricting the exotic matter to an arbitrarily thin region under
the condition that $b'(r)$ is close to unity near the throat. If
$b'(r_0)\approx 1$, then the embedding diagram will flare out very
slowly, so that, from Equation (\ref{r0}), $r_0$ may be made
arbitrarily small. Nevertheless, using the form functions
specified in~\cite{Morris:1988cz}, we will consider that $|b'-1|\approx
1$. Using the above value of $|\zeta-1|=10^{-10}$, then from
Equation (\ref{r0}) we find $r_0 \simeq 10^4\,{\rm m}$.

To determine the traversal time of the total trip, suppose that
the traveler accelerates at $g_{\oplus}$ halfway to the throat,
then decelerates at the same rate coming to rest at the \mbox{throat
\cite{Morris:1988cz,Kuhfittig:2002ur}}. For simplicity, we shall consider $b'
\approx 1$ near the throat, so the  wormhole will flare out
very slowly. Therefore, from Equation (\ref{r0}), the throat will be
relatively small, so  we may assume that $r_0 \approx 0$, as
in~\cite{Kuhfittig:2002ur}. Consider that a space station is situated
just outside the junction surface at $a$, from which the intrepid
traveler will start their journey. Thus, the total time of the
trip will be approximately $t_{\rm total} \simeq 4
\sqrt{2a/g_{\oplus}} \approx 1800 \,{\rm s} = 30\,{\rm min}$. The
maximum velocity attained by the traveler will be approximately
$v_{{\rm max}} \simeq 4.5\,{\rm km/s}$.

\section{Conclusions}\label{secconclusion}

Traversable wormhole spacetimes typically require the presence of exotic matter, i.e., material that violates the NEC. However, by introducing a junction surface, it is possible to confine this exotic matter to a finite region, matching the interior wormhole geometry to an exterior, asymptotically well-behaved spacetime. As a result, the exotic matter can, in principle, be restricted to an arbitrarily small volume.
The volume integral quantifier 
provides a means to evaluate the total amount of energy condition-violating matter present~\cite{Visser:2003yf}. Using this formalism, explicit wormhole solutions have been constructed where appropriate choices of the interior metric and matching radius yield configurations supported by arbitrarily small quantities of matter that violate the averaged null energy condition \mbox{(ANEC)~\cite{Visser:2003yf}}.

Nevertheless, violations of the NEC, its averaged form (ANEC), and the WEC are inherent features of traversable wormholes. These arise directly from the flaring-out condition at the throat, which entails the violation of the NEC.
To minimize the presence of exotic matter, as mentioned above, one can consider models where these violations are confined to a narrow region near the throat, $r_0 \leq r < a$, while the thin shell at the junction surface $\Sigma$ may, in principle, satisfy the standard energy conditions. In the limit $a \rightarrow r_0$, the exotic matter region becomes arbitrarily small. In this construction, the stress--energy tensor on the shell can be interpreted as describing quasi-normal matter, meaning that it obeys both the NEC and the WEC~\cite{Visser:1995cc,Hawking:1973uf,Visser:2003yf}.

It is worth noting that in this context, analyzing the dominant energy condition (DEC) becomes largely redundant. The DEC, which imposes that energy flows along causal (timelike or null) worldlines and demands that the energy density dominate over the pressures, is a stricter constraint than both the WEC and NEC. Since the NEC is already violated in all traversable wormhole geometries, this automatically implies the breakdown of both the WEC and DEC as well. Therefore, the violation of the DEC is not an independent or surprising feature in wormhole physics, and it is sufficient to focus the analysis on the NEC and WEC violations that are inherently tied to the wormhole's geometric structure.

In summary, in this work, we have explored the construction and physical viability of thin-shell wormholes influenced by a cosmological constant, employing the cut-and-paste technique. Our analysis revealed that the inclusion of a positive cosmological constant, as in the Schwarzschild–de Sitter geometry, notably enlarges the parameter space of stable configurations compared to the classic Schwarzschild case. Conversely, for a negative cosmological constant, corresponding to the Schwarzschild–anti-de Sitter background, the stability regions contract, indicating a more constrained and potentially less favorable environment for sustaining traversable wormholes.

To reduce the reliance on exotic matter, a generic feature in wormhole physics, we further constructed composite spacetimes by joining an interior wormhole geometry to an exterior vacuum solution at a finite junction surface. By examining the surface stresses, we identified domains where the weak and null energy conditions could be respected. A governing equation was derived for the radial pressure's behavior across the junction, adding further insight into the mechanical equilibrium of the configuration.

Finally, in evaluating traversability, we considered realistic constraints such as the proper throat and junction radii, traversal time, and allowable velocities. These estimates ensure that the constructed wormholes are not only theoretically consistent but also potentially traversable. Altogether, our findings broaden the framework for building physically plausible and stable wormholes, and demonstrate the crucial role a cosmological constant may play in shaping their viability within general relativity.

In conclusion, the thin-shell construction remains a powerful and versatile method for modeling wormholes, particularly in scenarios where one aims to minimize exotic matter. Its ability to clearly examine the properties of the junction surface makes it useful for theoretical investigations into the interplay between gravitation, energy conditions, and topology change. Looking ahead, future work could extend these models by incorporating more general matter fields, analyzing dynamical stability under time-dependent perturbations, or embedding them in modified theories of gravity. Such directions not only deepen our understanding of fundamental physics but also pave the way for exploring whether traversable wormholes might one day be realized within a consistent and physically plausible framework. 

\vspace{6pt} 



\funding{
	FSNL acknowledges support from the Funda\c{c}\~{a}o para a Ci\^{e}ncia e a Tecnologia (FCT) Scientific Employment Stimulus contract with reference CEECINST/00032/2018, and the research grants No. UID/FIS/04434/2020, No. PTDC/FIS-OUT/29048/2017 and No. CERN/FIS-PAR/0037/2019.
}

\conflictsofinterest{The authors declare no conflicts of interest.} 


\begin{adjustwidth}{-\extralength}{0cm}

\reftitle{References}

\PublishersNote{}
\end{adjustwidth}

\end{document}